\documentclass[10pt,conference]{IEEEtran}

\usepackage[utf8]{inputenc}

\usepackage[american]{babel}

\usepackage{csquotes}
\usepackage{lipsum} 
\usepackage{acronym}
\usepackage{graphicx}
\usepackage{caption}

\newtheorem{constraint}{Constraint}
\newtheorem{milpConstraint}{MILP-Constraint}
\newtheorem{objective}{Objective}
\newtheorem{milpObjective}{MILP-Objective}
\usepackage{gensymb}
\usepackage{amsmath}
\usepackage{amssymb}

\usepackage{algorithm}
\usepackage[noend]{algpseudocode}

\usepackage{booktabs} 
\usepackage{tabulary}
\usepackage{tabularx}
\usepackage{multirow}
\usepackage{colortbl}

\usepackage{mathtools}
\usepackage[hidelinks]{hyperref}
\usepackage{cleveref}

\crefname{constraint}{constraint}{constraints}
\Crefname{constraint}{Constraint}{Constraints}
\Crefname{objective}{Objective}{Objective}

\crefname{milpConstraint}{MILP-constraint}{MILP-constraints}
\Crefname{milpConstraint}{MILP-Constraint}{MILP-Constraints}
\Crefname{milpObjective}{MILP-Objective}{MILP-Objectives}

\usepackage[numbers]{natbib}

\usepackage{tikz}
\usetikzlibrary{arrows}
\usetikzlibrary{automata}
\usetikzlibrary{calc}
\usetikzlibrary{fit}
\usetikzlibrary{positioning}
\usetikzlibrary{graphs}
\usetikzlibrary{matrix}
\usepackage{pgfplots}

\pgfplotsset{select coords between index/.style 2 args={
    x filter/.code={
        \ifnum\coordindex<#1\fi
        \ifnum\coordindex>#2\fi
    }
}}

\newcommand{\squeeze}{\vspace{-0mm}}
\let\oldcaption\caption
\renewcommand{\caption}[1]{\oldcaption{\small #1 \normalsize}}

\usepackage{siunitx}
\DeclareSIUnit[]\cycles{\text{cycles}}
\DeclareSIUnit[]\bit{\text{b}}
\DeclareSIUnit[]\sample{\text{S}}
\DeclareSIUnit{\dbm}{%
 \text{dBm}%
}
\begin{document}
\newacro{fa}[FA]{Factory Automation}
\newacro{mac}[MAC]{Medium Access Control}
\newacro{uav}[UAV]{Unmanned Aerial Vehicle}
\newacro{ibr}[IBR]{Institut f\"ur Betriebssysteme und Rechnerverbund}
\newacro{phy}[PHY]{Physical Layer}
\newacro{qos}[QoS]{Quality of Service}
\newacro{cpn}[CPN]{Cyber-Physical Network}
\newacro{cps}[CPS]{Cyber-Physical System}
\newacro{5g}[5G]{fifth generation}
\newacro{koi}[KOI]{Koordinierte Industriekommunikation}
\newacro{lte}[LTE]{Long-Term Evolution}

\newacro{rss}[RSS]{Received Signal Strength}
\newacro{ntp}[NTP]{Network Time Protocol}
\newacro{lan}[LAN]{Local Area Network}
\newacro{iot}[IoT]{Internet of Things}
\newacro{tdoa}[TDoA]{Time Difference of Arrival}
\newacro{toa}[ToA]{Time of Arrival}
\newacro{pmu}[PMU]{Phase Measurement Unit}
\newacro{wsn}[WSN]{Wireless Sensor Network}
\newacro{pll}[PLL]{Phase Locked Loop}
\newacro{fft}[FFT]{Fast Fourier Transformation}
\newacro{ism}[ISM]{Industrial Scientific Medical}
\newacro{wlan}[WLAN]{Wireless Local Area Network}
\newacro{us}[US]{ultrasound}
\newacro{gps}[GPS]{Global Positioning System}
\newacro{dgps}[DGPS]{Differential Global Positioning System}
\newacro{rtb}[RTB]{Ranging Toolbox}
\newacro{dqf}[DQF]{Distance Quality Factor}
\newacro{ar}[AR]{Active Reflector}
\newacro{ppv}[PPV]{Positive Predictive Value}
\newacro{psd}[PSD]{Power Spectral Density}
\newacro{rf}[RF]{Radio Frequency}
\newacro{cots}[COTS]{Commercial Off-The-Shelf}
\newacro{uwb}[UWB]{Ultra Wide Band}
\newacro{oqpsk}[O-QPSK]{Offset Quadrature Phase-Shift Keying}
\newacro{mac}[MAC]{Medium Access Control}
\newacro{tdma}[TDMA]{Time Division Multiple Access}
\newacro{lwb}[LWB]{Low-power Wireless Bus}
\newacro{hart}[HART]{Highway Addressable Remote Transducer}
\newacro{ocari}[OCARI]{Optimization of Communication for Ad-hoc Reliable Industrial Networks}
\newacro{whart}[WirelessHART]{WirelessHART}
\newacro{isa}[ISA]{International Society of Automation}
\newacro{wisa}[WISA]{Wireless Interface for Sensors and Actuators}
\newacro{imu}[IMU]{Inertia Measurement Unit}
\newacro{dcs}[DCS]{Distributed Control System}
\newacro{iwsn}[IWSN]{Industrial Wireless Sensors Network}
\newacro{wsanfa}[WSAN-FA]{Wireless Sensor/Actuator Network for Factory Automation}
\newacro{gts}[GTS]{Guaranteed Time Slot}
\newacro{mars}[MARS]{Mobility-Aware Real-Time Scheduling for Low-Power Wireless Networks}
\newacro{mrt}[MoRT]{Mobile Real Time}
\newacro{frtos}[FreeRTOS]{Free Real Time Operating System}
\newacro{ptp}[PTP]{Precision Time Protocol}
\newacro{nic}[NIC]{Network Interface Controller}
\newacro{isr}[ISR]{Interrupt Service Routine}
\newacro{cc}[CC]{Complex Channel}
\newacro{cdf}[CDF]{Cumulative Distribution Function}
\newacro{stm32}[STM32]{STM32 discovery kit with STM32F407VG MCU}
\newacro{uart}[UART]{Universal Asynchronous Receiver Transmitter}
\newacro{std}[SD]{Standard Deviation}
\newacro{ci}[CI]{Constructive Interference}
\newacro{ct}[CT]{Concurrent Transmission}
\newacro{st}[ST]{Single Transmission}
\newacro{lqi}[LQI]{Link Quality Indicator}
\newacro{per}[PER]{Packet Error Rate}
\newacro{rxstart}[\textit{rx\_start}]{reception start interrupt}
\newacro{rxend}[\textit{rx\_end}]{reception end interrupt}
\newacro{txend}[\textit{tx\_end}]{transmission end interrupt}
\newacro{txstart}[\textit{tx\_start}]{transmission start signal}
\newacro{pps}[PPS]{Pulse Per Second}
\newacro{spi}[SPI]{Serial Peripheral Interface}
\newacro{sdr}[SDR]{Software Defined Radio}
\newacro{dsss}[DSSS]{Direct Sequence Spread Spectrum}
\newacro{tc}[TC]{Task Cluster}
\newacro{milp}[MILP]{Mixed Integer Linear Programming}
\newacro{maxage}[\textit{maxAge}]{maximum age}
\newacro{maxjitter}[\textit{maxJitter}]{maximum jitter}
\newacro{lcm}[LCM]{least common multiple}
\newacro{cllf}[C-LLF]{Conflict-aware Least Laxity First}
\newacro{dag}[DAG]{Directed Acyclic Graphs}


%
\title{Adaptive Real-Time Scheduling for Cooperative \aclp{cps}}
%
%


\author{\IEEEauthorblockN{Georg von Zengen, Jingjing Yu and Lars C. Wolf}
\IEEEauthorblockA{Institute of Operating Systems and Computer Networks\\
Technische Universit\"at Braunschweig\\
Email: [vonzengen$\mid$jyu$\mid$wolf]@ibr.cs.tu-bs.de}
}

\maketitle

\thispagestyle{plain}
\pagestyle{plain}

\begin{abstract}
\acsp{cps} are widely used in all sorts of applications ranging from industrial automation to search-and-rescue.
So far, in these applications they work either isolated with a high mobility or operate in a static networks setup.
If mobile \acsp{cps} work cooperatively, it is in applications with relaxed real-time requirements.
To enable such cooperation also in hard real-time applications we present a scheduling approach that is able to adapt real-time schedules to the changes that happen in mobile networks.
We present a \acl{milp}-model and a heuristic to generate schedules for those networks.
One of the key challenges is that running applications must not be interrupted while the schedule is adapted.
Therefore, the scheduling has to consider the delay and jitter boundaries, given by the application, while generating the adapted schedule.
\end{abstract}

\begin{IEEEkeywords}
Scheduling algorithms, Cyber-Physical Systems, Real-time systems, Wireless networks
\end{IEEEkeywords}

%
\IEEEpeerreviewmaketitle

\section{Introduction}

\label{sec:intro}
In traditional real-time systems a schedule was calculated once and used until the system got another taskset.
To switch the taskset of such systems they were stopped completely and started with the new schedule.
However, stopping the whole application is not an option in systems that have to perform their tasks continuously.
An example for such systems are mobile robots that cooperate to carry a work piece.
If a third robots needs to join the group of the carrying robots to weld a second piece to the carried one, the schedules need to be changed without stopping the application on the carrying robots.
Stopping the application might lead to dropping the work peace as it is now longer kept in balance by a feedback loop closed between the robots.
We call these groups of robots \acp{tc}, as they perform a common task cooperatively.
To overcome this challenge, mechanisms are needed to generate schedules that introduce the changes, needed to adapt to the new situation, without harming the real-time constraints of running tasks.
In this paper we introduce two different mechanisms to overcome these challenges.
The first one is a \ac{milp}-model that is able the schedule such \acp{tc} and also to merge their schedules.
The second one is a heuristic algorithm with the same capabilities to be used on embedded devices, as it is less computational complex.
This algorithm is based on a hypothesis we introduce and test prior to the discussion of the algorithm.
This paper is structured as follows, in \Cref{sec:AllocSchedProblem} we introduce the problem in more detail and discuss our assumptions.
In \Cref{sec:sched-constr-object} we present the constraints a scheduling algorithm need to fulfill to solve the priorly stated problem.
\Cref{sec:AllocSchedRelated} discusses related work from different areas that handles scheduling of problems similar to ours.
In \Cref{sec:AllocSchedMilp} we introduce our \ac{milp}-model to schedule mobile \acp{cps} and to adapt schedules to changes.
Afterward we evaluate the computational complexity in \Cref{sec:eval-comp-compl}.
As solving \ac{milp}-models is a rather complex task, we introduce and test a hypothesis on which schedules are better adaptable than others in \Cref{sec:adaptibilityHypothesis}.
To calculate schedules on embedded devices in a more predictable time we discuss a heuristic algorithm in \Cref{sec:AllocSchedEuch}.
This heuristic algorithm is evaluated in \Cref{sec:scheduleAlgoEvaluation}.
\Cref{sec:alloc-concl} concludes the paper and its contributions.

\section{Problem Statement}
\label{sec:AllocSchedProblem}
For our scheduling we assume that a distributed real-time system is a wireless network consisting of several nodes.
Each node is capable of executing different tasks but only one of them per time-slot.
Nodes can only communicate in a half-duplex manner on one channel in one time-slot but may switch channels between two consecutive time-slots.
The whole network on the other hand is able to utilize multiple channels at the same time to transfer data in disjoint sets of nodes.
The problem of scheduling data transfers over multiple channels can be mapped to the problem of scheduling computation tasks to multiple processors.
\citeauthor*{coffman1972}~\cite{coffman1972} and \citeauthor*{du1989}~\cite{du1989} showed that multiprocessor scheduling of non-preemptive task is NP-Hard.

All nodes are able to use the same set of interference free channels.
Each channel has the same characteristics for all nodes in the \ac{tc}.
Further, we assume that all channels have the same characteristics.
We also assume all nodes participating in a \ac{tc} to be in each others transmission range, thus all communication is single-hop.

Several tasks from a job, the tasks of a job might be executed on several nodes and have dependencies between each other.
Each job has a period which is inherited to the tasks.
All jobs of a \ac{tc} form the set $\omega$.
Jobs might share common tasks, these tasks have the shortest period of all jobs they are participating in.
Besides the period ($P_i$), each task $T_{i}$ has a maximum jitter ($J_i$) that describes how many time-slots a task might move between two consecutive executions.
Additionally, each task $T_{i}$ has a maximum age ($d_{ij}$) that describes how many time-slots the task might be executed before a task ($T_j$) depending on it.
The tasks $T_i$ depends on are called its dependencies and are stored in $\Gamma_i$.
The matrix $D$, with dimensions $[|\tau| \times |\tau|]$, stores how many time-slots a task $T_i$ is allowed to be scheduled before task $T_j$.
All tasks in a \ac{tc} form the set $\tau$.
The execution time of each task is assumed to be at most one time-slot.

As we assume that the job of a system fails if one task is not executed in time we do not implement priorities in our scheduling.

All tasks of all jobs in a system form a directed graph without circles.
This graph might have several entry tasks and leaf tasks.
Each job in this graph forms a so-called path that might have several entry tasks but only one leaf task.
Entry tasks are tasks without dependencies, they form the set $E$.
Leaf tasks are tasks without dependent tasks, all leaf tasks form the set $L$. 

The number of time-slots in which the schedule is not repeated is called Hyperperiod\,($H$).
It is defined as the least common multiple of all periods of jobs in $\omega$.

We differentiate between time-slot and slot, a time-slot is a certain portion of time that has a defined start time and end time, all time-slots are of the same length.
A time-slot can inhabit multiple slots, as a slot is a time-slot on a certain network resource.
Therefore, a time-slot consists of as many slots as interference free network resources are available.
An example for interference free network resources are the channels a communication standard defines.
In a system with only one communication standard that defines three channels, each time-slot would consist of three slots.

Two tasks are called intersecting if they have common communication partners.
This is the case if they have a direct dependency to each other, they are depending on the same task or they have the same depending task.
Intersections between tasks are stored in the matrix $I$ with the dimensions $[|\tau| \times |\tau|]$.

\begin{figure}[ht]
  \centering
  \squeeze
  \includegraphics[width=.7\linewidth]{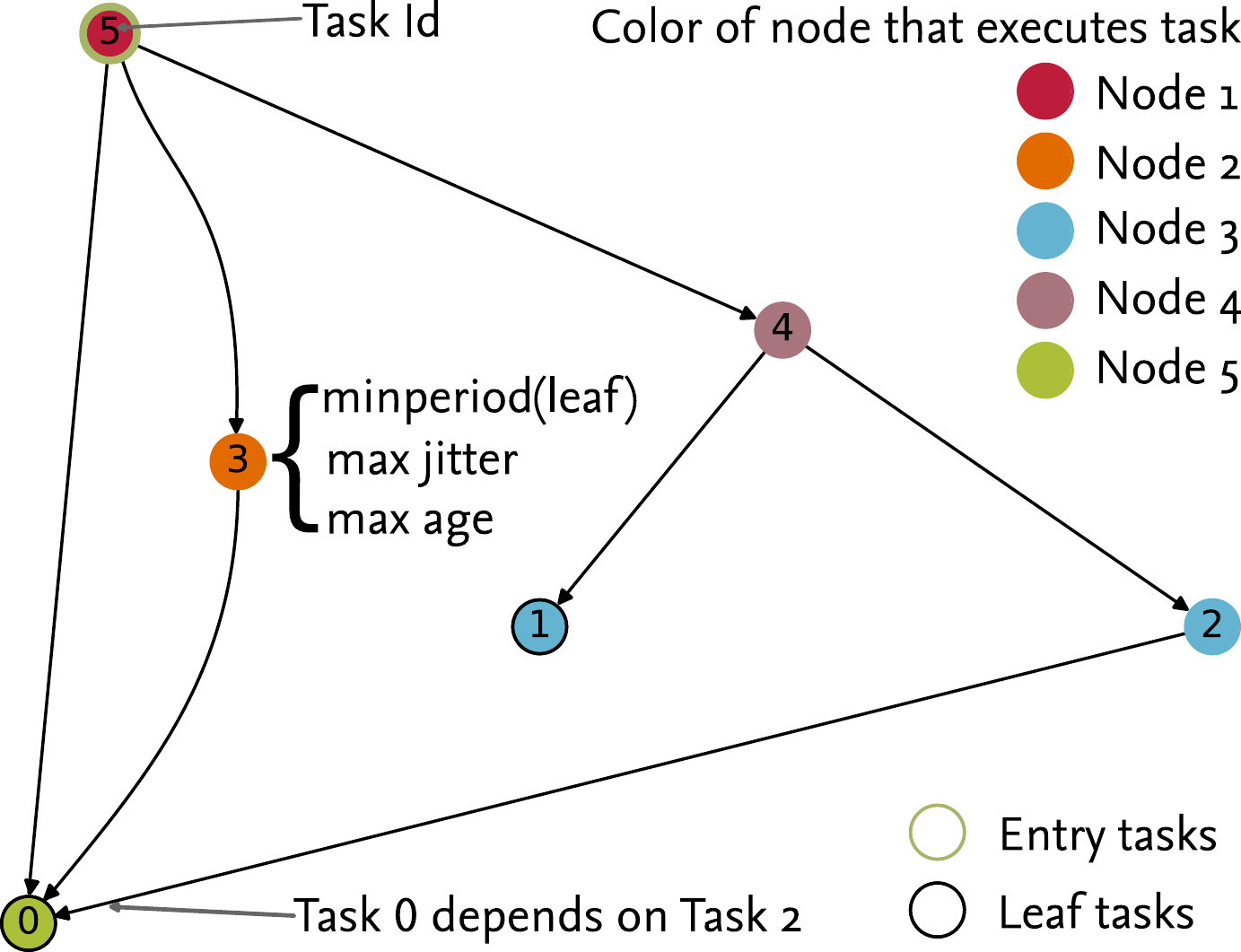}
  \caption{Example dependency graph with two jobs consisting in total of six task executed on a network with five nodes. \label{fig:AllocSchedGraph} \squeeze}
\end{figure}

\Cref{fig:AllocSchedGraph} depicts an example dependency graph that consists of two jobs.
The first job has the entry task with id 5 and the leaf task with id $0$, called job\,$0$.
It consists of the tasks: 5, 4, 3, 2  and $0$.
The second job has the entry task with id 5 and the leaf task with id 1, called job\,1.
Consisting of the tasks: 5, 4 and 1.
Both of them share the common tasks 5 and 4.
The graph is executed on a network with five nodes, the color of each task shows on which node it needs to be executed on.
The arrows between tasks describe the dependencies, where the task the arrow is pointing to is depending on the task the arrow is pointing from, e.g. task\,$0$ is depending on task\,5. 

\section{Scheduling Constraints and Objectives}
\label{sec:sched-constr-object}
This section discusses the constraints the scheduling has to fulfill in order to generate valid schedules.
\begin{constraint}
  \label{textcon:1}
  It is not allowed that two or more tasks share the same slot.
\end{constraint}
Sharing slots would lead to transmission interference and loss of data, therefore each task needs its own slot.

\begin{constraint}
  \label{textcon:2}
  Tasks with a common participating node must not be executed in the same time-slot
\end{constraint}
\Cref{textcon:2} prohibits the scheduler from scheduling two tasks at the same time that have the same node either receiving or transmitting data.
In our example from \Cref{fig:AllocSchedGraph} there are different tasks that can not be scheduled at the same time: task\,1 and task\,2 as they are executed at the same node, task\,2 and task\,3 as both transmit data that is needed by task\,$0$.
By prohibiting these tasks to be scheduled at the same time, this constraint ensures that the communication can be handled by half-duplex transceiver.

\begin{constraint}
  \label{textcon:3}
  All dependencies of a task must be scheduled before the depending task.
\end{constraint}
Referring to our example in \Cref{fig:AllocSchedGraph} this constraint ensures that, e.g., task\,5 is scheduled before task\,$0$, task\,3 and task\,4. 
To be able to schedule task\,$0$ it is necessary that task\,5, task\,3 and task\,2 are scheduled before task\,$0$.

\begin{constraint}
  \label{textcon:4}
  Each dependency of a task must be scheduled no more than its maximal age before the task.
\end{constraint}
As most data has an age at which it becomes less usable to the requesting task, the providing task or dependency must be scheduled less than this age before the requesting task.
E.g., if the maximal age of data provided by task\,3 is ten time-slots, task\,3 must not be scheduled more than ten slots before task\,$0$. 

\begin{constraint}
  \label{textcon:5}
  All depending tasks in one job must use the same execution of a common dependency.
\end{constraint}
To ensure that tasks in one job use the same state of the system, it is necessary that tasks in one job that depend on the same task are scheduled after the same execution of that task.
In our example: the job with entry task task\,5 and leaf task task\,$0$ is formed by the task ids 5, 4, 3, 2 and $0$.
\Cref{textcon:5} defines that task\,5 must not be scheduled between the tasks 4, 3, 2 and $0$.

\begin{constraint}
  \label{textcon:6}
  Each leaf task must be scheduled once in its period.
\end{constraint}
The different jobs might have different periods in which they have to be scheduled in, the leaf task of each jobs might have a different period.
Therefore, some of these periods might differ form the hyperperiod, which is the \ac{lcm} of all periods.
To fulfill the requirements of all jobs, the leaf might have be scheduled multiple times in one schedule.
Together the \Cref{textcon:3,textcon:4,textcon:5,textcon:6} ensure that each job is executed the right amount of times per hyperperiod and all tasks in the jobs are executed in the right order.

\begin{constraint}
  \label{textcon:7}
  Two consecutive periods of the same task must not exceed the defined jitter bound for this task.
\end{constraint}
As described in \Cref{sec:AllocSchedProblem}, each task has a jitter bound that must not be exceeded.
Therefore, \Cref{textcon:7} ensures that the period of a task does not vary more than its jitter bound.
E.g., the scheduled period of a task with the defined period of five slots and a maximal jitter of two slots could be decreased to three slots or increased to seven slots.
But two consecutive periods of that task may not vary more than two slots, thus, a period change from three slots to seven is not allowed.

A scheduler that enforces all the \Crefrange{textcon:1}{textcon:7} will generate schedules applicable to a system described in \Cref{sec:intro}.
To be able to adapt to topology changes, the scheduler needs to follow one more constraint:
\begin{constraint}
  \label{textcon:8}
  The difference between the last period in the old schedule and the first period in the new schedule of a task must not exceed the defined jitter bound for this task.
\end{constraint}
If this constraint is followed the network can switch its schedule to the new one without breaking any real-time constraints.

To make the operation of a network reliable the schedule stability should be maximized.
This is especially important when existing schedules need to be adapted to changes in the topology or taskset of a network.
Increasing the schedule stability reduces the complexity of switching the schedules and therefore reduces the probability of failures while switching.
Therefore, we formulate the general objective for the scheduling as following:
\begin{objective}
  \label{obj:general}
  Time-slot changes between periods of tasks should be minimized.
\end{objective}

\section{Related Work}
\label{sec:AllocSchedRelated}
Real-time scheduling is a vast topic, especially if the scope is widened to real-time multiprocessor scheduling.
To keep this section of reasonable size we focus on the most applicable related work.

\ac{cllf} was proposed by \citeauthor{saifullah2010} to schedule \acl{whart} networks\cite{saifullah2010}.
It is designed for wireless real-time networks with changing topologies.
The priority of a transmission is determined by the time between it is released and its deadline and the number of conflicting transmissions in this time.
The highest priority is given to the task with the highest number of conflicting transmissions and the shortest time between release and deadline.
\ac{cllf} need the release time of each transmission prior to scheduling, this is not possible in systems where the release time of a transmission depends on a task that has dependencies.
As it is unknown when a dependency is scheduled, the release time of the dependent transmission is also unknown.
Therefore, \ac{cllf} is not applicable for the problem described in \Cref{sec:AllocSchedProblem}. 

Another application of wireless real-time networks with changing topologies are \ac{lte} networks.
\citeauthor{shakkottai2001} introduced an algorithm to schedule a mixture of real-time and non-real-time traffic in \ac{lte} networks\cite{shakkottai2001}.
At each time-slot the algorithm calculates which of the packets in the transmission queue has the shortest deadline and schedules this packet into the slot at a certain channel.
This done by the eNode-B for each transmission time interval which consists of multiple time-slots.
The schedule determined in this manner is only valid for the down-link traffic from the eNode-B to the user equipment.
This technique is only possible, as the eNode-B buffers all down-link traffic.
In a network such as in \Cref{sec:AllocSchedProblem} there is not one central instance as the eNode-B that buffers all the traffic.

\citeauthor{wang2019} propose a two staged approach to adaptive scheduling in train communication networks\cite{wang2019}.
These networks are often time-triggered and need to handle rapid topology changes in cases where two trains are coupled or decoupled.
The first stage is the offline scheduling, it generates schedules for the whole train.
The second stage is called online stage, it derives the schedule for the parts of the train during the coupling and decoupling process.
A two staged design seems applicable to our system as we also expect rapid topology changes when two \acp{tc} need to be merged.
On the other hand the characteristics of the network described by \citeauthor{wang2019} are in great contrast to our application.
The authors describe train networks as strictly hierarchical multicluster networks with wired connections.
Further, the approach does not have concept of dependencies between different data flows.\\

A novel approach on how to close feedback control over wireless links is proposed by \citeauthor{baumann2019}\cite{baumann2019}.
They propose a system that reduces the communication between different parts of a distributed feedback controller.
The reduction is achieved by a co-design called control-guided communication.
Control-guided means that the controller tells the communication part of the system its communication demands ahead of time.
The communication demand is decreased by a controller that estimates values in between communication.
To benefit from the decreased communication demands the schedule needs to reschedule the communication frequently.
Therefore, \citeauthor{baumann2019} choose an online scheduling approach.
Like the other approaches this approach also lacks concept of dependencies between tasks.

As the approaches discussed above do lack the ability to handle dependencies the scope is widened to scheduling in operations research.
In assembly lines dependencies are very common and therefore scheduling approaches in this domain need to handle them from the beginning.
The algorithm proposed by \citeauthor{hu1961} forms \acp{dag} from a given sets of tasks\cite{hu1961}.
In these \acp{dag} each node represents a task and each directed edge represents a dependency.
The author assumes an assembly line with a number of equally skilled workers, each of this workers is able to fulfill one task at a time.
The task have a predefined order in which they have to be fulfilled, the goal is to find the sequence of tasks for each worker that needs the shortest time to complete all tasks of the given set
To clarify why this is applicable to the challenges stated in \Cref{sec:AllocSchedProblem} let the equally skilled worker be equally good channels and the predefined order gives a set of dependencies between tasks.
This gives a model of a system which is quite close to the one we depicted above.
However, a difference between our use case and assembly line scheduling is, that there is no concept of tasks that can not be executed at the same time because of common child tasks.
Further, the tasks in \citeauthor{hu1961}'s model have no deadline and the goal is to finish as fast as possible, in contrast our goal is to meet the deadline of all task.

Each of the scheduling concepts mentioned addresses a part of the problem stated in \Cref{sec:AllocSchedProblem}.
However, each one is missing some key features needed to fulfill the requirements for scheduling algorithms as they are needed in this work. 
Some of the solutions used in the following two approaches are inspired by the discussed concepts.

\section{\acl{milp} Approach}
\label{sec:AllocSchedMilp}
In this section we introduce an approach to solve the problem described in \Cref{sec:AllocSchedProblem} based on a \ac{milp} model.
First we introduce the model for a decision problem that generates valid schedules regarding to the assumptions discussed in \Cref{sec:AllocSchedProblem}.
Afterwards, we describe the constraints needed by the model to generate valid schedules.
Then we introduce the objective that minimizes the jitter.

To ease the modeling we extend the schedule $S$ by a third dimension which is the dimension of tasks in $\tau$, the resulting Matrix is called $A$.
Thus, the three dimensions of $A$ are, (i) the considered task in $\tau$, (ii) the channels and (iii) the time-slots.
$A$ has the dimensions $|\tau| \times M \times H$ and is defined as followed:
\squeeze
\begin{equation}
\forall a_{Tct} \in A : a \in \{0,1\}
\end{equation}
Where $T$ is the task, $c$ is the channel and $t$ is the time-slot.
\begin{equation}
  a_{Tct}=
  \begin{cases*}
    1,& task $T$ is scheduled in channel $c$ \\
      & at time-slot $t$ \\
    0,& task $T$ is not scheduled in channel $c$\\
      & at time-slot $t$
  \end{cases*}
\end{equation}

Following the constraints of the \ac{milp}-model are discussed in detail.

\squeeze
\begin{milpConstraint}
  \label{con:1}
   Every slot (timeslot and channel) must at most have one scheduled task
\vspace{-1mm}
  $$ \sum_{T \in \tau} a_{Tct} \leq 1 \quad $$ 
  $$\forall c \in \mathbb{N}: 1 \leq c \leq M,\;\forall t \in \mathbb{N}: 1 \leq t \leq H $$ 
\end{milpConstraint}
\vspace{-1mm}
\Cref{con:1} guarantees that there are no direct collisions between tasks in the same time-slot and the same channel and therefore implements \Cref{textcon:1}.

\begin{milpConstraint}
  \label{con:2}
   Tasks with common participating node must not be executed in the same time-slot
  $$ (\iota_{UT} \times \sum_{c=1}^{M} a_{Tct}) + (\iota_{UT} \times \sum_{c=1}^{M} a_{Uct})\leq1 $$
  $$\quad \forall t \in \mathbb{N}: 1 \leq t \leq H,\; \forall U,T \in \tau$$
\end{milpConstraint}
As task that have an intersection in their sets of nodes they are communication with must not be scheduled in one time-slot, \Cref{con:2} takes the sum of all scheduled distributions of the interfering tasks $T$ and $U$ in a time-slot $t$ over all channels $1 \leq c \leq M$.
This sum has to be less or equal one for all time-slots and all interfering pairs of tasks.
Whether tasks $T$ and $U$ are interfering is determined from matrix $I$  at position $UT$. 
It is the implementation of \Cref{textcon:2}.
The matrix $I$ is defined by \Cref{eq:defI} and \Cref{eq:defIota}.
\begin{equation}
  \label{eq:defI}
\forall \iota_{UT} \in I : \iota \in \{0,1\}
\end{equation}
Where $T$ and $U$ are task in $\tau$ .
\begin{equation}
  \label{eq:defIota}
  \iota_{UT}=
  \begin{cases*}
    1,& tasks $T$, $U$ have common nodes \\
    0,& tasks $T$, $U$ do not have common nodes 
  \end{cases*}
\end{equation}
\begin{milpConstraint}
\label{con:3}
All dependencies $U$ of task $T$ must be scheduled before $T$ within the minimum of $P_i$ or $t-d_{U}$ time-slots   
$$ \sum_{i=max(1; t-d_{U}; \lfloor t/P_T \rfloor\times P_T)}^{max(1; t)} \sum_{c=1}^{M} a_{Uci} \geq \sum_{c=1}^{M}a_{Tct} $$ 
$$\forall t \in \mathbb{N}: 1 \leq t \leq H,\; \forall U \in \Gamma_T,\; \forall T \in \tau $$
\end{milpConstraint}
\Cref{con:3} ensures that all dependencies $\Gamma_T$ to a task $T$ are executed at least as often as the dependent task $T$.
It sums up all scheduled executions of a dependency $U$ in the $P_i$ slots and channels before $T$ is scheduled and ensures that this sum is larger than the sum of the scheduled executions of $T$ in time-slot $t$.
This guarantees that \Cref{textcon:3,textcon:4} are enforced.
\begin{milpConstraint}
  \label{con:4}
  Each leaf task must be scheduled once per its period
  $$ \sum_{t=max(1; (p-1)\times P_T}^{p\times P_T} ~~ \sum_{c=1}^{M} a_{Tct} = 1 $$
  $$\forall p \in \mathbb{N}: 1 \leq p \leq \frac{H}{P_T},\; \forall T \in \tau$$
\end{milpConstraint}
To ensure that each job is executed once in its period (\Cref{textcon:6}), \Cref{con:4} sums up all executions of task $T$ during all possible periods and ensures the sum is always one for all tasks.
\begin{milpConstraint}
  \label{con:5}
  Execution of a $T$ must be scheduled in its jitter bounds
  $$ \sum_{i=max(1;t-P_T-J_T)}^{max(1;t-P_T+J_T)} \sum_{c=1}^{M}a_{Tci} \geq \sum_{c=1}^{M}a_{Tct}$$ 
  $$\forall t \in \mathbb{N}: 1 \leq t \leq H,\; \forall T \in \tau $$
\end{milpConstraint}
As a task $T$ scheduled outside its jitter bound $J_T$ could harm the operation of the system (\Cref{textcon:7}),
\Cref{con:5} prohibits that.
Therefore it checks whether there is an execution of $T$ scheduled in the jitter bound $\pm J_T$ one period $P_T$ before the current execution. 
The following \Cref{con:8,con:10,con:11} ensure that all tasks of one job use the same execution of a common dependency (\Cref{textcon:5}).
Referring to the graphs in \Cref{fig:AllocSchedGraph} they ensure that task\,$0$, task\,3 and task\,4 all use the same execution of task\,5.
That means, task\,5 must not be scheduled between task\,$0$, task\,3 and task\,4.
This might happen, as jobs can share certain tasks and the periods of some dependencies might be smaller than the period of the job and the depending task. 
As an example, let job\,$0$ have a period of ten slots and job\,1 a period of five slots.
With the constraints \Crefrange{con:1}{con:5} there is nothing that prohibits to schedule the tasks\,4 and 5 a second time in between the scheduled executions of task\,3 and $0$.
Doing so could cause task\,$0$ to operate on different data than task\,3.


To cope with this we need to introduce a few more variables.
Let  $\Omega_{T_eT_l}$ be the set of tasks between the entry task $T_e$ of a job and its leaf task $T_l$ with the same period as $T_e$.

For the example above $\Omega_{50}$ would consist of the tasks 3, 2, 0.

On contrast let $\breve\Omega_{T_eT_l}$ be the tasks of the job that have a shorter period than $T_e$.
Again, for the example that would mean $\breve\Omega_{50}$ consists of the tasks 5 and 4.

\begin{milpConstraint}
  \label{con:8}
  $\rho_{T_eT_l}$ is the sum of all execution of all Tasks in $\Omega_{T_eT_l}$, $\breve\rho_{T_eT_l}$ is the sum of all execution of all Tasks in $\breve\Omega_{T_eT_l}$
  $$\rho_{T_eT_l} = \sum_{T \in \Omega_{T_eT_l} } \sum_{c=1}^{M} \sum_{i=t}^{min(i+P_{T_l}, H)}a_{Tci} $$ 
  $$\forall t \in \mathbb{N}: 1 \leq t \leq H, \forall T_e \in E,\; \forall T_l \in L  $$
  $$\breve\rho_{T_eT_l} = \sum_{T \in \breve\Omega_{T_eT_l}} \sum_{c=1}^{M} \sum_{i=t}^{min(i+P_{T_l}, H)}a_{Tci} $$ 
  $$\forall t \in \mathbb{N}: 1 \leq t \leq H, \forall T_e \in E,\; \forall T_l \in L  $$
\end{milpConstraint}
\Cref{con:8} is only needed to give $\rho_{T_eT_l}$ and $\breve\rho_{T_eT_l}$ a value that indicates how many tasks of $\Omega_{T_eT_l}$ and $\breve\Omega_{T_eT_l}$ are scheduled in one period of $T_l$.

\begin{milpConstraint}
  \label{con:10}
  If the complete path or non of its tasks is scheduled $\dot{\rho}_{T_eT_l}$ and $\dot{\breve\rho}_{T_eT_l}$ are 1. If only parts are scheduled $\dot{\rho}_{T_eT_l}$ and $\dot{\breve\rho}_{T_eT_l}$ are $0$.
  $$\rho_{T_eT_l} = |\Omega_{T_eT_l}| \times \dot{\rho}_{T_eT_l} \quad \forall T_e \in E,\; \forall T_l \in L$$
  $$\breve\rho_{T_eT_l} = |\breve\Omega_{T_eT_l}| \times \dot{\breve\rho}_{T_eT_l} \quad \forall T_e \in E,\; \forall T_l \in L$$
\end{milpConstraint}
$\rho_{T_eT_l}$ and $\breve\rho_{T_eT_l}$ are used in \Cref{con:10} to determine whether all task in $\Omega_{T_eT_l}$ or respectively $\breve\Omega_{T_eT_l}$ are scheduled.
That is necessary as tasks in $\breve\Omega_{T_eT_l}$ might be scheduled independently but tasks in $\Omega_{T_eT_l}$ must not be scheduled without the tasks in $\breve\Omega_{T_eT_l}$.
This in ensured by \Cref{con:11}

\begin{milpConstraint}
  \label{con:11}
  The $\breve\Omega_{T_eT_l}$ might be scheduled alone but $\Omega_{T_eT_l}$ must not
  $$ \dot{\breve\rho}_{T_eT_l} \geq \dot{\rho}_{T_eT_l} \quad \forall T_e \in E,\; \forall T_l \in L$$
\end{milpConstraint}

Together the \Crefrange{con:1}{con:11} define a model for the decision problem.
This model is able to generate valid schedules in $A$.
\subsection{Objectives}
\label{sec:objectives}
The objective of our model is to minimize the jitter between executions of a task.
Minimizing the jitter in feedback loops is one of more obvious optimizations for a scheduling.
This is due to the fact, that jitter leads to a bigger error in timing of the control task and therefore to larger error in the controlled process.

Minimizing the jitter in the model described above is challenging, as it lacks a concept of how many slot are between different executions of a task.
To mitigate this, we minimize the number of tasks that change their relative time-slot in different periods of their job.
With this optimization we additionally increase the schedule stability.
The schedule stability gives a measure how much a schedule changes between periods.
\begin{milpObjective} 
  \label{obj:1}
  Minimize the number of tasks changing slots between their periods 
  $$ Minimize: \frac{1}{N} \times \sum_{T \in \tau}~\sum_{t=1}^{H-P_T}\left(~\left\lvert\sum_{c=1}^{M}a_{Tct} - \sum_{c=1}^{M}a_{Tc(t+P_T)}\right\rvert\right)$$
\end{milpObjective} 

To evaluate the effectiveness of \Cref{obj:1} we scheduled the same set of 130000 tasksets once with \Cref{obj:1} and once without any objective.
These tasksets were randomly generated under certain constraints, we defined five different hyperperiod lengths (8, 12, 16, 25, 35), four different numbers of dependencies between the tasks (9, 12, 16, 24) and three different numbers of jobs (1 ,3 ,6), the number of tasks were eight or twelve.
\begin{figure}[ht]
  \centering
  \squeeze
    \centering\includegraphics[width=.8\linewidth]{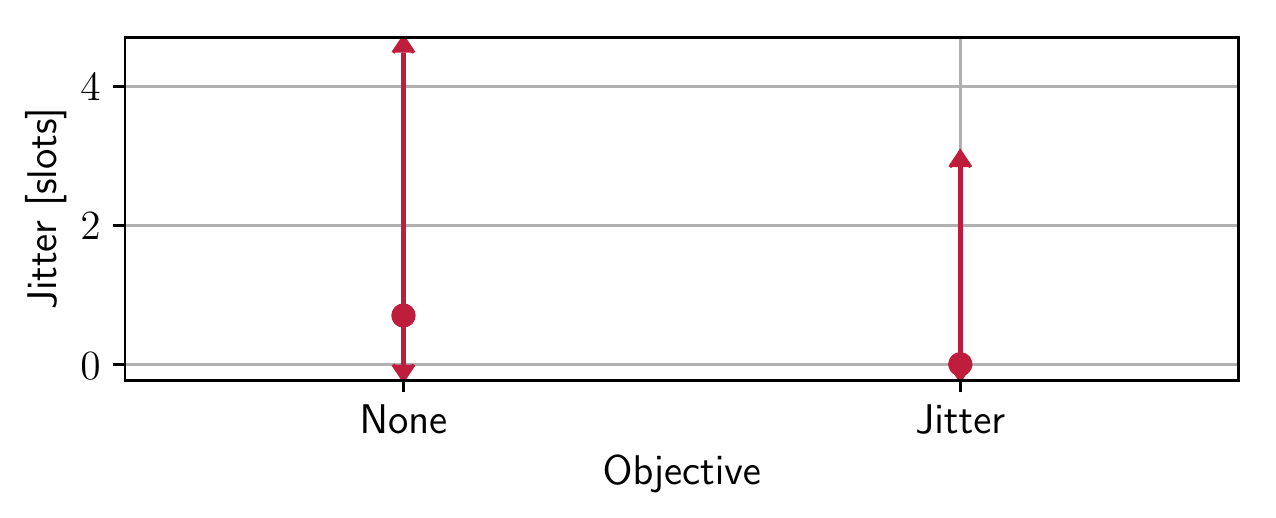}
  \caption{Mean, maximum and minimum jitter over all scheduled tasksets for the two objectives, ``None'' for no objective and ``Jitter'' for \Cref{obj:1}. \label{fig:AllocSchedJitter} \squeeze }
\end{figure}
In \Cref{fig:AllocSchedJitter} the jitter for the two different objectives is shown.
``None'' refers to the results scheduled without any ojective, ``Jitter'' refers to the jitter in the schedules optimizes to a minimum jitter by \Cref{obj:1}. 
The jitter shown in \Cref{fig:AllocSchedJitter} is calculated using \Cref{eq:AllocJitter}, it is the mean of the mean jitter of all tasks scheduled in the schedule.
\begin{equation}
  \label{eq:AllocJitter}
   jitter = \frac{ \sum_{T\in \tau}\frac{\sum_{i=1}^{H/P_T} (e_{Ti} - e_{T1}) \: mod \; P_T} {\sum_{t=1}^{H}\sum_{c=1}^{M}a_{tcT}}}{|\tau|} 
\end{equation}
As \Cref{fig:AllocSchedJitter} shows \Cref{obj:1} reduces the mean jitter to almost zero.

As an objective can not harm the performance, in terms of schedulablility, of our \ac{milp}-model, we use the \Cref{obj:1} for all further evaluations if not stated otherwise.

\subsection{Adapting Schedules}        
\label{subsec:adaptSchedules}
This section discusses one of the key contributions of this chapter, the adaptation of existing schedules to changes in the topology of the network or in the taskset.
The adaptation needs to be done without harming the real-time requirements of jobs which are present in the existing schedule.
\Cref{textcon:8} formulates this complex goal in a very brief way.

To achieve the goal of adapting schedules, several steps are needed in preparation.
First the tasksets of the old schedule and the new tasks need to be merged.
The new tasks can either be a second taskset of another \ac{tc} or tasks of new job added to the existing \ac{tc}.
In this work we focus on the first case, where two \acp{tc} need to be merged into one.
We consider this as the more complex case, as adding a new job is the same despite the fact, that the new job does not have the restrictions of an old schedule.
While joining the tasksets ($\tau_1$ and $\tau_2$), task-ids must be kept unique throughout the new taskset $\tau'$.
The new hyperperiod $H'$ is the \ac{lcm} of all periods in $\tau'$.

In the second step the two schedules ($A_1$ and $A_2$) are merged into one schedule $C$ that violates \Cref{textcon:1}.
Thus, tasks of both networks might share one time-slot on the same channel.
As $C$ is never to be executed, this does not cause any harm to the networks.
$C$ is used in \Cref{con:12} in addition to \Crefrange{con:1}{con:11} to generate the new, combined schedule $A'$ that respects the \Crefrange{textcon:1}{textcon:8}.
\begin{milpConstraint}
  \label{con:12}
    Timeslot of $task_T$  must not differ more than $J_T$  from $C$  to $A'$, changes of channels are ignored
    $$ \sum_{j=t-J_T}^{t+J_T}~\sum_{c=1}^{M}a'_{Tcj} ~\geq ~\sum_{c=1}^{M}c_{Tct} \quad \forall t \leq H',\; \forall T \in \tau'$$
\end{milpConstraint}

Together with \Cref{con:12} we introduce the new \Cref{obj:reschedule} that minimizes the amount of task that are shifted to other time-slots between $C$ and $A'$.
Thus, the schedule stability is maximized.
  \begin{milpObjective}
    \label{obj:reschedule}
    Minimize time-slot allocation changes from $C$ to $A'$
    $$ Maximize: \sum_{T \in \tau'}\sum_{t=1}^{H'} ( \sum_{c=1}^{M}a_{Tct}' \times   \sum_{c=1}^{M}c_{TcT})$$
  \end{milpObjective}
By multiplying the sum of all channel for a certain time-slot and a certain task in the new schedule with the sum of the channels of the same task and time slot in the combined schedule, we get one for each task that was not move and zero for each task that was moved.
As each task can at most be scheduled once per time-slot the result of this multiplication can only have the two values, one and zero.
By summing this result up over all tasks and time-slots, we get the number of the unchanged time-slot allocations.
As we maximize this value, the schedule stability is maximized and the jitter is minimized.

\section{\acl{milp}-model and its Complexity}
\label{sec:eval-comp-compl}
To calculate schedules that respect \Cref{textcon:1,textcon:8}, we implemented a \ac{milp}-model.
By solving this \ac{milp}-model with its \Cref{obj:general} we get a base line of optimal schedules.

As stated in \Cref{sec:AllocSchedProblem} multi-channel real-time scheduling is NP-Hard.
Therefore, it is important to evaluate whether the formulated \ac{milp}-Model can be solved in a reasonable time according to the use case.
Besides that we will also investigate wjhat parameters influence the time it takes to solve the \ac{milp}-Model.
As the first parameter we evaluated the number of slots in a hyperperiod.
We scheduled more than 160,000 different task-sets with five different hyperperiods: 8, 12, 16, 25 and 35 slots on an Intel Xeon W-2195 CPU.
These task-sets were randomly generated following guidelines in the hyperperiod length, number of dependencies, nodes, jobs, etc.
Form these parameter of the task-sets the hyperperiod length is the one that influences the computation time the most.
The time used to solve the model varies from 0.07\,s to more than 600\,s.
\Cref{fig:solvetime_slots} shows the \ac{cdf} of the five different hyperperiods.
The right graph shows only the range from zero to ten seconds of the left graph to show more details.
As expected, a longer hyperperiod leads to a longer solve-time, this is due to the larger solution space.
The dots in both graphs mark the maximum solve-time.
Even though, some task-sets need over 600\,s the majority is scheduled in less than 3\,s.
\begin{figure}[ht]
  \centering
 \squeeze 
  \includegraphics[width=.97\linewidth]{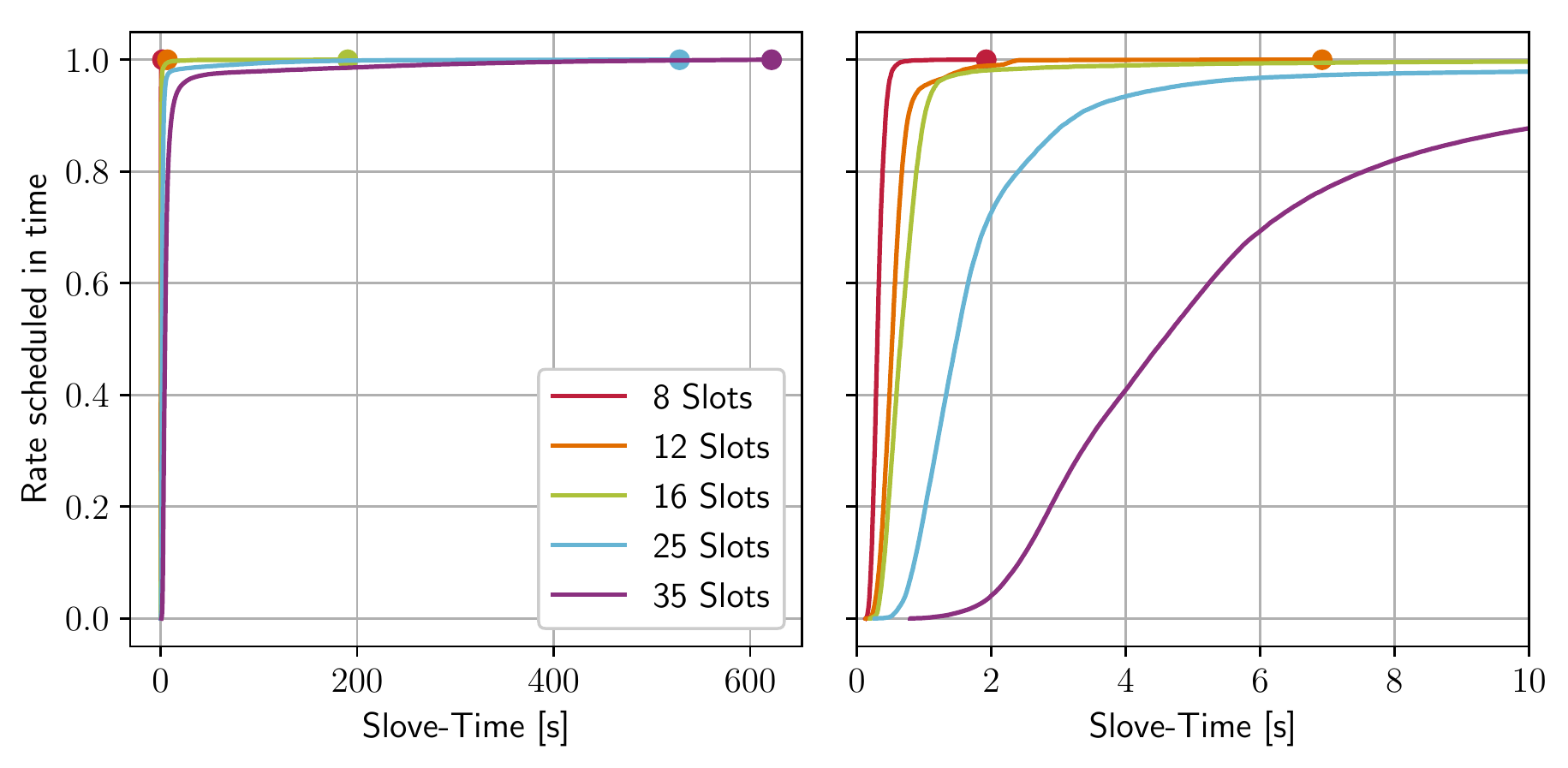}
  \caption{Impact of different hyperperiod length to the time needed to solve the schedule, shown as \ac{cdf} \label{fig:solvetime_slots}  \squeeze }
\end{figure}

\begin{figure}[hbt]
  \centering
  \includegraphics[width=.97\linewidth]{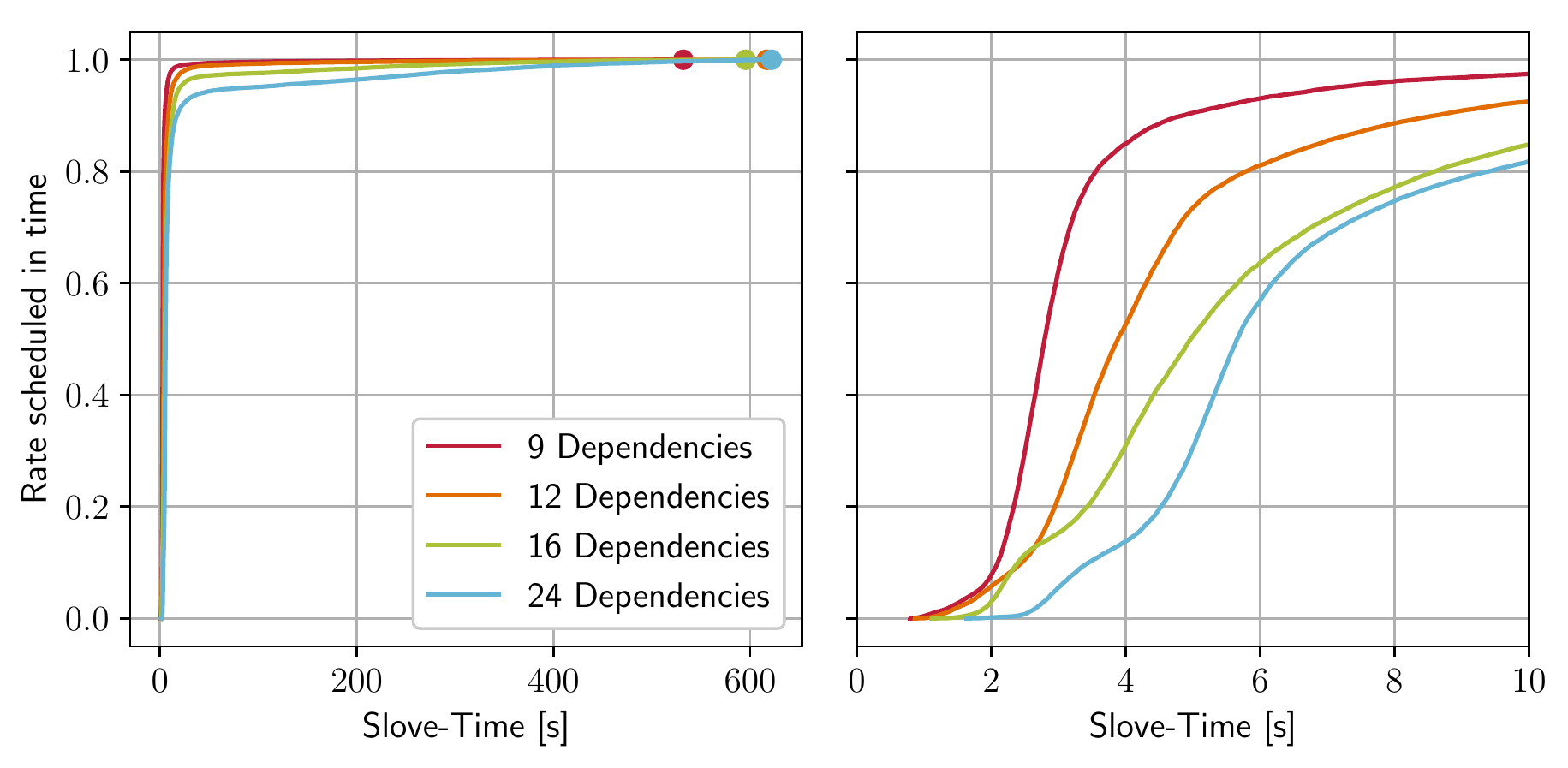}
  \caption{Impact of different numbers of dependencies to the time needed to solve the schedule, shown as \ac{cdf} \label{fig:solvetime_dependencies} }
\end{figure}

As a second parameter we evaluated the influence of the number of dependencies in a taskset.
To mitigate the influence of the hyperperiod length we only evaluate the solve-time of tasksets with 35 slots.
\Cref{fig:solvetime_dependencies} shows the \ac{cdf} for 9, 12, 16, and 24 dependencies, all together is show 76000 tasksets.
As the model gets more complex with more dependencies the solve-time increases as well.

\begin{figure}[hbt]
  \centering
  \includegraphics[width=.97\linewidth]{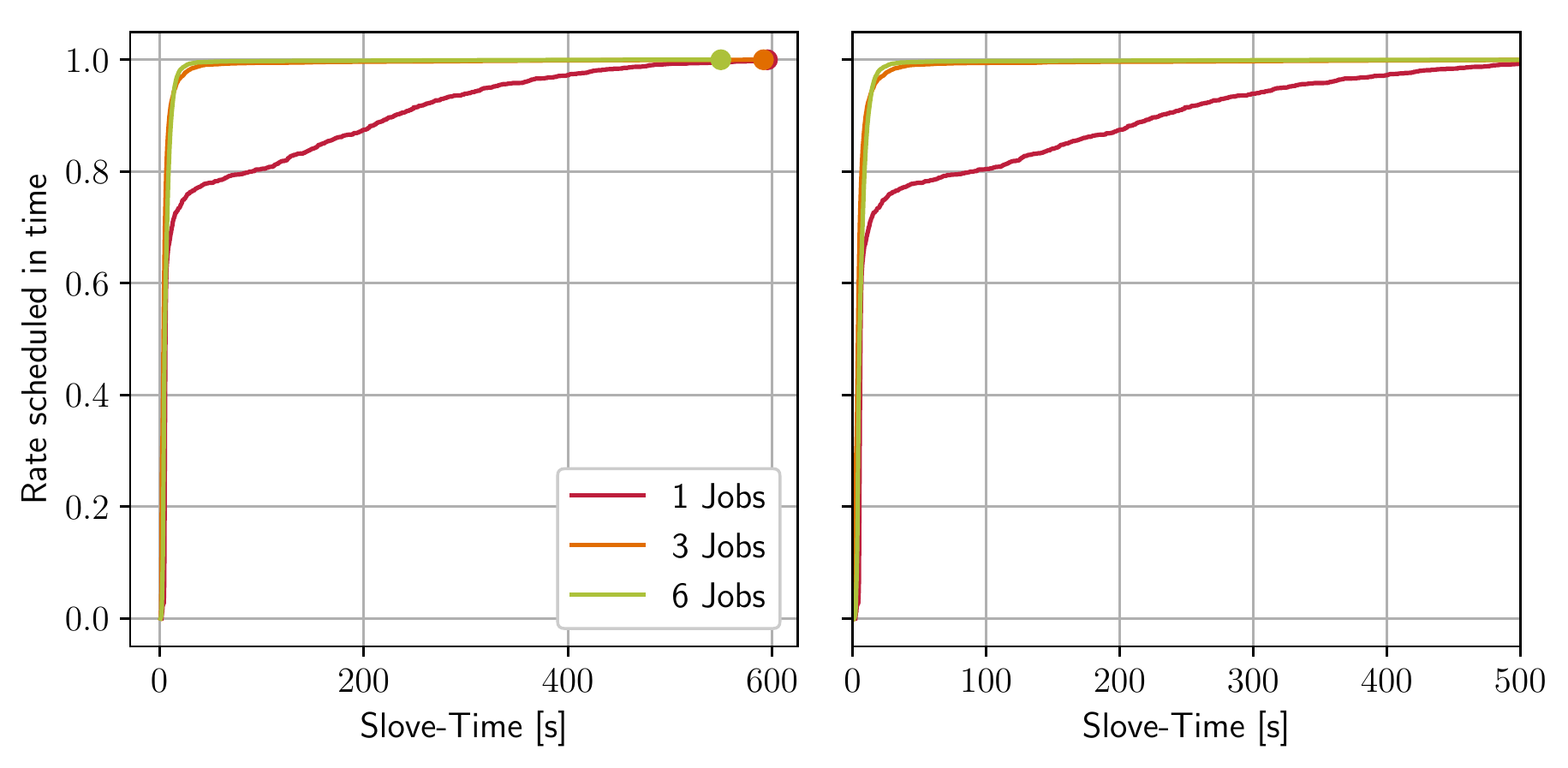}
  \caption{Impact of different numbers of jobs to the time needed to solve the schedule, shown as \ac{cdf} \label{fig:solvetime_jobs} }
\end{figure}

To determine how \acp{tc} with multiple jobs would be handled in contrast to \acp{tc} with just one job, we evaluate the solve-time of tasksets with a hyperperiod of 35 slots and 16 dependencies.
These roughly 20000 task sets have either 1, 3 or 6 jobs. 
\Cref{fig:solvetime_jobs} shows that a taskset with more complex jobs, more dependencies, takes longer to be scheduled as an easier one.

\subsection{Applicability to Embedded Devices}
As most \acp{cps} consist of embedded devices with far less computation power than our Intel Xeon W-2195 (fast CPU), we assume the solve-time to be higher.
To support this assumption, we used our 10 years old Intel Xeon E5520 (slow CPU) to schedule all tasksets with hyperperiod of 35.
The results for the median solve-time are not that different: \SI{5.5}{\second} for the slow CPU and \SI{4.5}{\second} for the fast CPU.
However, the mean and maximum solve-times differ a lot: the fast CPU needs a mean solve-time of \SI{11.7}{\second} and a maximum of \SI{621}{\second}, the slow CPU needs \SI{72.3}{\second} in mean and \SI{14332}{\second} at maximum.
This huge variety in the solve-times is a problem considering real-time applications.
Even if the scheduling has no hard time constraints, waiting up to  \SI{600}{\second} or even \SI{14000}{\second} for a schedule is unrealistic in most applications.
Therefore, a way to calculate schedules in a more predictable time is necessary.

In \Cref{sec:adaptibilityHypothesis} we formulate a hypothesis on the adaptability of schedules and validate it to get better insight on how to design an algorithm suitable for the described problem.

\section{Hypothesis on Adaptability of Schedules}
\label{sec:adaptibilityHypothesis}
In most scheduling applications it is preferable to schedule all tasks as dense as possible.
So the taskset can be scheduled more often in the same amount of time or the executing machines can sleep or take other jobs.
In a system where jobs have a defined period there is no need to schedule all task as dense as possible.
In contrast, it might have advantages to schedule tasks as sparse as possible.
That means, free slots are more uniformly distributed throughout the hyperperiod.

\begin{figure}[ht]
  \centering
  \includegraphics[width=.8\linewidth]{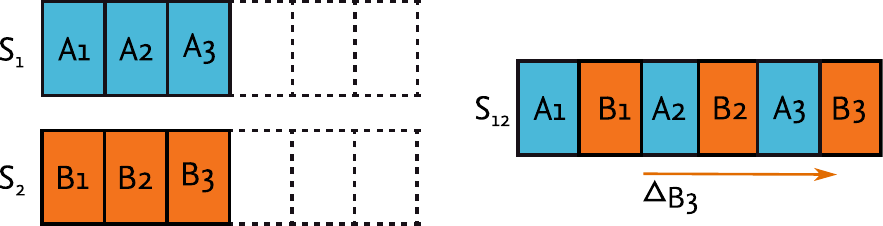}
  \caption{Example of dense schedules. \label{fig:denseSchedule} }
\end{figure}

This is especially advantageous if adaptations are taken into consideration.
Having free slots throughout the hyperperiod means that tasks in a merged schedule must not be shifted to time-slots as far as in a dense schedule.
\Cref{fig:denseSchedule} depicts two densely schedules $S_1$ and $S_2$, $S_1$ has the tasks $A_1$ to $A_3$ and $S_2$ the tasks $B_1$ $B_3$.
The merged schedule $S_{12}$ contains all tasks.
In this example all tasks are scheduled very densely in the first three of the six time-slots.
To merge $S_1$ and $S_2$ the tasks are shifted into other time-slots.
$B_3$ has to be shifted three time-slot, therefore $\Delta_{B_3}$ is 3.

\begin{figure}[ht]
  \centering
  \includegraphics[width=.8\linewidth]{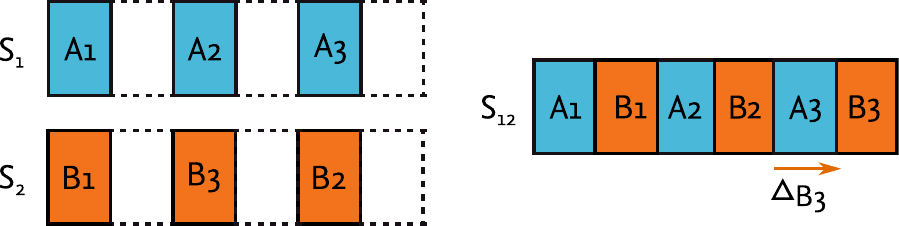}
  \caption{Example of sparse schedules. \label{fig:sparseSchedule} }
\end{figure}

\Cref{fig:sparseSchedule} on the other hand shows two sparse schedules that also contain three tasks in six time-slots.
In contrast to the schedules in \Cref{fig:denseSchedule}, this time the tasks are distributed equally over the six time-slots.
To merge the schedules all tasks of $S_2$ have to be shifted only by one time-slot, therefore $\Delta_{B_3}$ is 1.

\Cref{fig:denseSchedule,fig:sparseSchedule} show the two extreme cases but they illustrate why it might be a good idea to spread the tasks throughout the whole hyperperiod.
In the following we introduce our metric for the degree of distribution and investigate under which circumstances the hypothesis, that pairs of schedules which have a higher distribution are more likely to be combinable, is true.
\subsection{Task Distribution}
\squeeze
\begin{equation}
  \label{metric:distribution}
Distribution = \frac{\sum_{t=1}^{H} x_t}{\sum_{t=1}^{H}\sum_{c=1}^{M}\sum_{T\in \tau}a_{tcT}} 
\end{equation}
\Cref{metric:distribution} gives the distribution as a normalized function of number of used time-slots to unused time-slots divided by the number of all scheduled executions of all tasks.
Where $x_t$ indicates an unused time-slot following an used time-slot, as described in \Cref{eq:x_t}.
\begin{equation}
  \label{eq:x_t}
x_t  = \begin{cases*}
    1,& if time-slot $t-1$ is used and\\
      &time-slot $t$ is unused \\
    0,& if time-slot $t-1$ is unused \\
      &or time-slot $t$ is used 
  \end{cases*}
   \forall 1 \leq t \leq H
\end{equation}
A time-slot is called used if there is a task scheduled on at least one channel.

\subsection{Validity of the Hypothesis}
\begin{figure}[hbt]
  \centering
\squeeze
  \includegraphics[width=.9\linewidth]{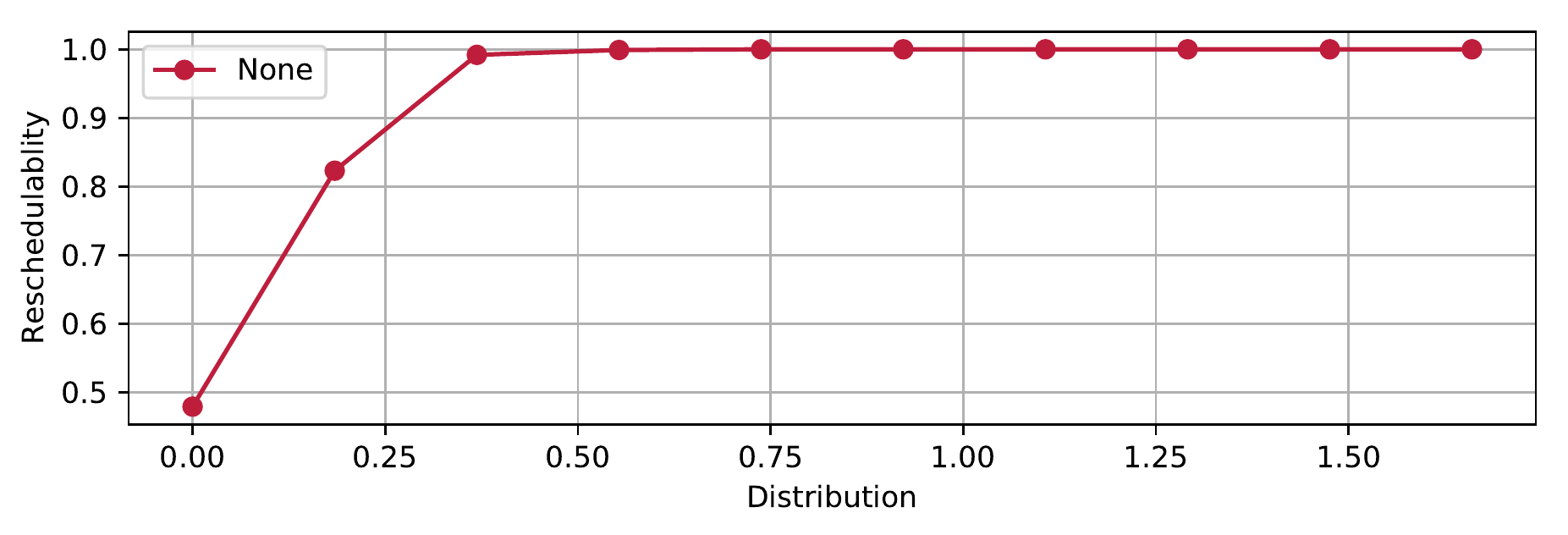}
  \caption{Reschedulablity over distribution \label{fig:reschedulable_distribution}  \squeeze}
\end{figure}

To validate whether the hypothesis is true we generated pairs of two schedules for two different tasksets of similar form, in terms of hyperperiod length, number of jobs, number of dependencies, etc.
The distribution of a pair lies between zero and two, as it is the sum of the distribution of both schedules.
These pairs were than rescheduled using the \ac{milp}-model.
\Cref{fig:reschedulable_distribution} shows that pairs with a higher distribution are much more likely to be schedulable than pairs with a lower distribution.

\section{Heuristic Approach}
\label{sec:AllocSchedEuch}
As we have shown in \Cref{sec:adaptibilityHypothesis} schedules with higher distribution can be merged more easily.
Therefore, we propose an algorithm that has three main goals, 1. maximize the distribution, 2. to minimize the jitter of tasks and 3. minimize the number of executions of each task.

In general the algorithm schedules a whole job before it starts to schedule the next job.
It starts with the job that has the longest path from the entry to the leaf task and continuous with the next longest job until all jobs are scheduled.
We call the job that is currently scheduled active.

Within a job the scheduler starts with the leaf task and places it in the latest possible slot, this way the number of slots to schedule the rest of the jobs is maximized.
The slot to schedule the leaf task for the $k$-th subperiod is calculated by utilizing \Cref{eq:leafSlot}.
A subperiod is one period of a job, its length is always a divider of the hyperperiod.
\begin{equation}
  t_{l,k} = k \times P_l
  \label{eq:leafSlot}
\end{equation}
As slots might be occupied by tasks which were scheduled prior to the current one, we introduce two solutions to choose another slot in which the current task is scheduled.
One of them chooses to use other channels before moving to other time-slots.
This should lead to a small jitter, as more tasks are scheduled in the calculated time-slot.
The other one prefers to use other time-slots first and only uses different channels if no time-slot in the allowed jitter bound is free.
This should be beneficial if \acp{tc} need to be merged.
In this case the \acp{tc} can simply use another channel and most conflicts are resolved.
These solutions are described in more detail in \Cref{subsec:timeFristShifting} and \Cref{subsec:channelFirstShifting}.
After a free slot is found, this slot in the two dimensional array $S$ is marked occupied with the task.
The dimensions of $S$ are given by the number of channels $M$ and length of the hyperperiod $H$.

After the leaf task is scheduled the algorithm uses the so-called backward equation to find the slot for the dependencies of the leaf task, this equation is explained in more detail in \Cref{subsec:backward-equation}.
The backward equation maximizes the distance of the execution between the scheduled task and other dependencies but also between the dependency and its dependencies.
This process is repeated until all entry tasks of a job are scheduled.

As it might occur that several jobs have common tasks, the backward equation would schedule a common task twice although an already scheduled execution of that task could be used to schedule the rest of the job.
Therefore, the algorithm searches the schedule whether there are tasks of the active job scheduled in a certain range of slots.
If such a task is found, the algorithm uses this execution and schedules the rest of the active job in the direction from that execution to the leaf task using the so-called forward equations, described in \Cref{subsec:forward-equation}.

While the scheduler is following the dependencies of tasks it might face tasks that have multiple dependencies or multiple tasks that depend on a task.
In both cases the scheduler needs to decide in which order these task should be scheduled.
We have two different approaches to this issue: the first one orders the task by ascending maximal age.
We call this approach \textit{age first}.
The idea is to schedule the task first that have a smaller range of time-slots in which they can be scheduled.

The second approach is, to sort the tasks by ascending maximal jitter.
This guarantees that the tasks with the hardest jitter constraints are scheduled first.
We call this approach \textit{jitter first}.
Both approaches will be compared in the evaluation \Cref{sec:scheduleAlgoEvaluation}.


\subsection{Backward Equation}
\label{subsec:backward-equation}
The Backward \Cref{eq:Backward} gives a time-slot in which the task should be scheduled, based on the execution time-slot of the tasks that are depending on the task to be scheduled.
$k$ gives the subperiod which is currently to be scheduled, $T_c$ is the child (dependent task) of a task and $T_p$ is the parent (dependency) of a task. 

\Cref{eq:Backward} needs a special case for the first subperiod $k=1$.
This is simply because in this case there is no prior execution of this tasks, so its time-slot can not be taken into the equation.
\small{
\begin{equation}
t_{p,k} = 
	\begin{cases}
		t_{c,k}-min( \lfloor \dfrac{(t_{c,k}-1)}{|\omega_i|-\delta_{p}} \rfloor, d_{p}), &\text{if }\ k=1 \\[10pt]
		t_{c,k}-min(\lfloor \dfrac{t_{c,k}-t_{c,k-1}-1}{|\omega_i|-\delta_{p}} \rfloor, d_{p}), &\text{if }\ 2 \leq k \leq H/P_{l}
	\end{cases}
\label{eq:Backward}
\end{equation}
}
With $|\omega_i|-\delta_{p}$ we calculate how many tasks of the job we have to schedule until the leaf task is reached.
The divided $t_{c,k}-1$ or $t_{c,k}-t_{c,k-1}-1$ gives the number of slots left in the subperiod.
The division of both gives a time-slot for the parent that has the maximal distance to the child but still leaves enough time-slots to schedule its parents.
As the time-slot calculated this way might have a larger distance to the child than the maximal age of the parent the minimum of the division and the maximal age is taken.

If the calculated time-slot is occupied, an alternative is selected applying the solutions from \Cref{subsec:timeFristShifting} or \Cref{subsec:channelFirstShifting}.

\subsection{Forward Equation}
\label{subsec:forward-equation}
Jobs in a taskset might have tasks in common, an example is depicted in \Cref{fig:AllocSchedGraph}.
In the example the jobs with the leaf task $T_1$ and $T_2$ have the common task $T_{4}$.
This common task is called $T_{com}$.
This fact can be used to reduce the total number of task executions in a schedule.
To do so we need to find these common tasks in the schedule and decide whether the found execution of such a common task fulfills the timing requirements of the child task that is to be scheduled based on the common task.
Therefore, the algorithm defines a range of time-slots in which the execution of a common task has to be located in order to use it.
The lower bound of this range $Lower_{com,k}$ for the $k$-th subperiod of $T_{com}$ is defined as the maximum of three values: first, the execution time-slot of its child task in the $k-1$-th subperiod.
This guarantees that the order of executions is not altered for prior subperiods.
The second value is the time-slot of $T_{com}$s execution in the $k-1$-th subperiod on which the period of the leaf task of the active job $P_l$ is added.
To be able to use extra time-slots the maximal jitter $J_{com}$ is subtracted.
The last value is the end of the current subperiod $k \times P_l$, to make use of allowed jitter $J_l$ is subtracted as well as the sum of all maximal ages of the tasks from the common task to the leaf task $\sum_{i=com}^{l-1}d_i$.
Thus, $Lower_{com,k}$ is defined as follows:
\small{
\begin{equation}
  \begin{aligned}
    Lower_{com,k} = max(&t_{c,k-1} +1, t_{com,k-1}+P_{l}-J_{com}, \\
    &k\times P_{l} - J_{l}-\sum_{i=com}^{l-1} d_{i})
  \label{eq:lowerSearchbox}
  \end{aligned}
\end{equation}
}

The upper bound of the range is defined as the minimum two values:  $t_{l,k} - \delta_{com}$ where $t_{l,k} = k \times P_l + J_L$, which gives the last slot far enough for the end of the subperiod to schedule all tasks between $T_{com}$ and the leaf task, and where $\delta_{com}$ is the distance $T_{com}$ and $T_l$.
The second value is the period of the leaf task $P_l$ and maximal jitter of $T_{com}$ added to the time-slot of the last execution of $T_{com}$.
\small{
\begin{equation}
Upper_{com,k} = min(t_{l,k} - \delta_{com}, t_{com, k-1}+P_{l}+J_{com})
 \label{eq:upperSearchbox}
\end{equation}
}


If an execution of $T_{com}$ was found in the range defined by \Cref{eq:upperSearchbox} and \Cref{eq:lowerSearchbox} the scheduler uses \Cref{eq:genForward} to calculate the time-slots the child $T_c$ of $T_{com}$ should be executed.
\begin{equation}
\small{
\begin{aligned}
&t_{c,k} = \\
&\begin{cases}
  t_{p,k}+min(\lfloor \dfrac{min(k\times  P_{l}, t_{p,k+1})-t_{p,k}}{\delta_{p}} \rfloor, d_{p})	&\textit{if}\ T_{p}=T_{com} \\
  & and \ k< H/P_{l}\\
t_{p,k}+min(\lfloor \dfrac{k\times  P_{l}-t_{p,k}}{\delta_{p}} \rfloor, d_{p})	&\textit{else}\ \\
\end{cases}
\end{aligned}
\label{eq:genForward}
}
\end{equation}
As \Cref{eq:genForward} describes the $k$-th execution of $T_c$ is scheduled after the $k$-th execution of $T_p$ but before the $k+1$-th execution of $T_p$.
Therefore, the execution order defined in the task set is respected.
The other limiting factor for the forward equation is the maximal age $d_p$ of $T_p$, the child must be scheduled before $d_p$, otherwise the data produced by $T_p$ is useless.
As for the backward equation it is possible that $t_{c,k}$ is already occupied, strategies to handle such situations are described in \Cref{subsec:timeFristShifting} and \Cref{subsec:channelFirstShifting}.
\subsection{Time First Shifting}
\label{subsec:timeFristShifting}
In the sections above we described how the algorithm determines in which time-slot a task should be scheduled.
If this slot is occupied on one channel the algorithm needs to find another slot either on another channel or at another time.
This section describes a solution to this challenge, that tries to find another time-slot before it uses other channels.

As the algorithm normally schedules the tasks from the leaf to the entry task and thus from right to left in the example, the algorithm first tries to put $T_i$ into $t_{i,k}+1$ on the same channel.
This is the time-slot next to $t_{i,k}$ on the right, by going right first the algorithm leaves potentially more space where the yet unscheduled tasks need to be scheduled.
If $t_{i,k}+1$ would be occupied as well the algorithm would go to $t_{i,k}-1$ and not $t_{i,k}+2$ to minimize the jitter.
After all slots in the jitter bound of $T_i$ are tested and found occupied the algorithm would test $t_{i,k}$ at another channel and repeat the same search pattern if it is occupied as well.
Another reason not to take $t_{i,k}$ on the second channel would be that $T_j$ and $T_i$ are interfering tasks, in this case this slot would be handled as occupied. 

In this mode the algorithm tries to fit all tasks on one channel.
This is done under the hypothesis that two schedules that use one channel primarily are easier to merge, as one of the schedules could be shifted to another channel and most of the conflicts would be resolved.

\subsection{Channel First Shifting}
\label{subsec:channelFirstShifting}
This mode of the algorithm solves the same issue as the one described in \Cref{subsec:timeFristShifting} but by using all available channels before shifting the execution of $T_i$ in time.
After all channels, in the example two, are found occupied for $t_{i,k}$ the algorithm would try then all channels at $t_{i,k}+1$.
This mode has the advantages that it minimizes the jitter in the schedule and that it potentially leaves more time-slots entirely empty.
Having time-slots empty on all channels might be of advantage when two schedules need to be merged, as tasks can be shifted there to make room for tasks that can not be shifted due to stricter jitter bounds. 

In \Cref{sec:scheduleAlgoEvaluation} we will evaluate the proposed shifting mechanisms.

\subsection{Schedule adaption}
\label{subsec:scheduleAdaption}

To achieve the goal of adapting schedules, several steps are needed in preparation.
First the tasksets of the old schedule and the new tasks need to be merged.
The new tasks can either be a second taskset of another \ac{tc} or tasks of new job added to the existing \ac{tc}.
In this work we focus on the first case, where two \acp{tc} need to be merged into one.
We consider this as the more complex case, as adding a new job is the same despite the fact, that the new job does not have the restrictions of an old schedule.
While joining the tasksets ($\tau_1$ and $\tau_2$), task-ids must be kept unique throughout the new taskset $\tau'$.
The new hyperperiod $H'$ is the \ac{lcm} of all periods in $\tau'$.
After the tasksets are merged, the same algorithm is used to generate the new schedule.

The result of the backward \Cref{eq:Backward} has no dependencies to tasks outside of the active job, therefore it will not changes unless the job is changed which results in a new job.
Thus, the backward equation does not harm the real-time requirements while adapting schedules.

The search range definitions \Cref{eq:lowerSearchbox} and \Cref{eq:upperSearchbox} for $k=1$ sets the upper search bound to $k\times P_{l} + J_{l} - \delta_{com}$, and the lower bound to $|\omega_i| - \delta_{com}$.
As the results are not effected by any task outside the job, it will not harm real-time requirements during adaption.
For all subperiods where $k>1$, the search box is limited by the previous subperiod, that means that no subperiod with $k>1$ could harm the real-time requirements of a task if the first subperiod did not harm them.
The same is true for the forward \Cref{eq:genForward} as it is not effected by tasks outside the job.

As discussed the calculations of the time-slots can not give results outside the jitter bound and the shifting methods described in \Cref{subsec:timeFristShifting} and \Cref{subsec:channelFirstShifting} do not shift further than the allowed jitter.
Thus, no task can be effected by a jitter more than the allowed one during schedule changes.
Therefore, the same algorithm can be used to schedule a \ac{tc} initially and to adapt its schedule to new topologies or tasksets.

\section{ Evaluation}
\label{sec:scheduleAlgoEvaluation}
In this section we evaluate the algorithm discussed in \Cref{sec:AllocSchedEuch} and compare it with the results of the \ac{milp}-model.
First we show that the presented algorithm has a much more predictable execution time than the \ac{milp}-model.
Afterwards we compare the different approaches in terms of jitter and the percentage of scheduleable tasksets.
In \Cref{tb:modes} we list the short names used to distinguish the different modes of the algorithm in this section. 
\vspace{-2mm}
\begin{table}[hbt!]
\caption{ The four different modes of the scheduling algorithm}
\centering
\begin{tabular}{c c c}
\textbf{Approach} 			& \textbf{Sort tasks on same level by}	& \textbf{Mode}	\\
\hline
\multirow{2}{*}{Time First Shifting }	&  Age First					& (0, 0)		\\
						& Jitter First					& (0, 1)		\\
\hline
\multirow{2}{*}{Channel First Shifting}	& Age First					& (1, 0) 		\\
						& Jitter First				& (1, 1)		\\
\end{tabular}
\label{tb:modes}
 \squeeze
\end{table} \squeeze
        
\subsection{Computational Complexity Comparison}
\label{sec:eval-compl-comp}
To compare the computational complexity of the milp-model and our algorithm, we scheduled more than 160,000 different task-sets with five different hyperperiods: 8, 12, 16, 25 and 35 slots on an Intel Xeon W-2195 CPU.
These task-sets were randomly generated following guidelines in the hyperperiod length, number of dependencies, nodes, jobs, etc.
Form these parameter of the task-sets the hyperperiod length is the one that influences the computation time the most.
The time used to solve the model varies from 0.07\,s to more than 600\,s.
\Cref{fig:solvetime_slots} shows the \ac{cdf} of the five different hyperperiods.
The right graph shows only the range from zero to ten seconds of the left graph to show more details.
As expected, a longer hyperperiod leads to a longer solve-time, this is due to the larger solution space.
The dots in both graphs mark the maximum solve-time.
Even though, some task-sets need over 600\,s the majority is scheduled in less than 3\,s.
\begin{figure}[ht]
  \centering
 \squeeze 
  \includegraphics[width=.97\linewidth]{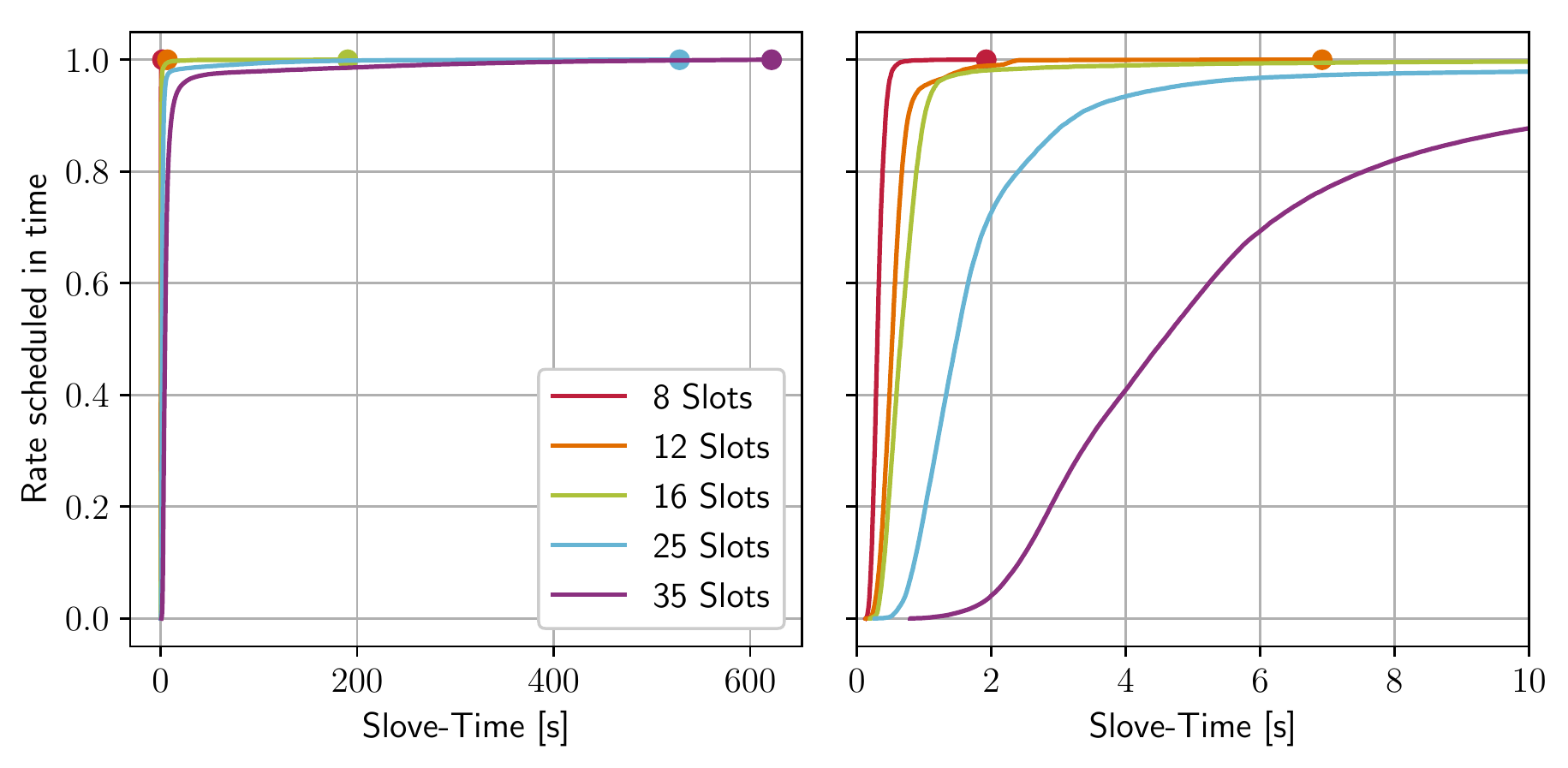}
  \caption{Impact of different hyperperiod length to the time needed to solve the schedule, shown as \ac{cdf} \label{fig:solvetime_slots}  \squeeze }
\end{figure}

\begin{figure}[hbt!]
\centering
\squeeze
\includegraphics[width=0.9\linewidth]{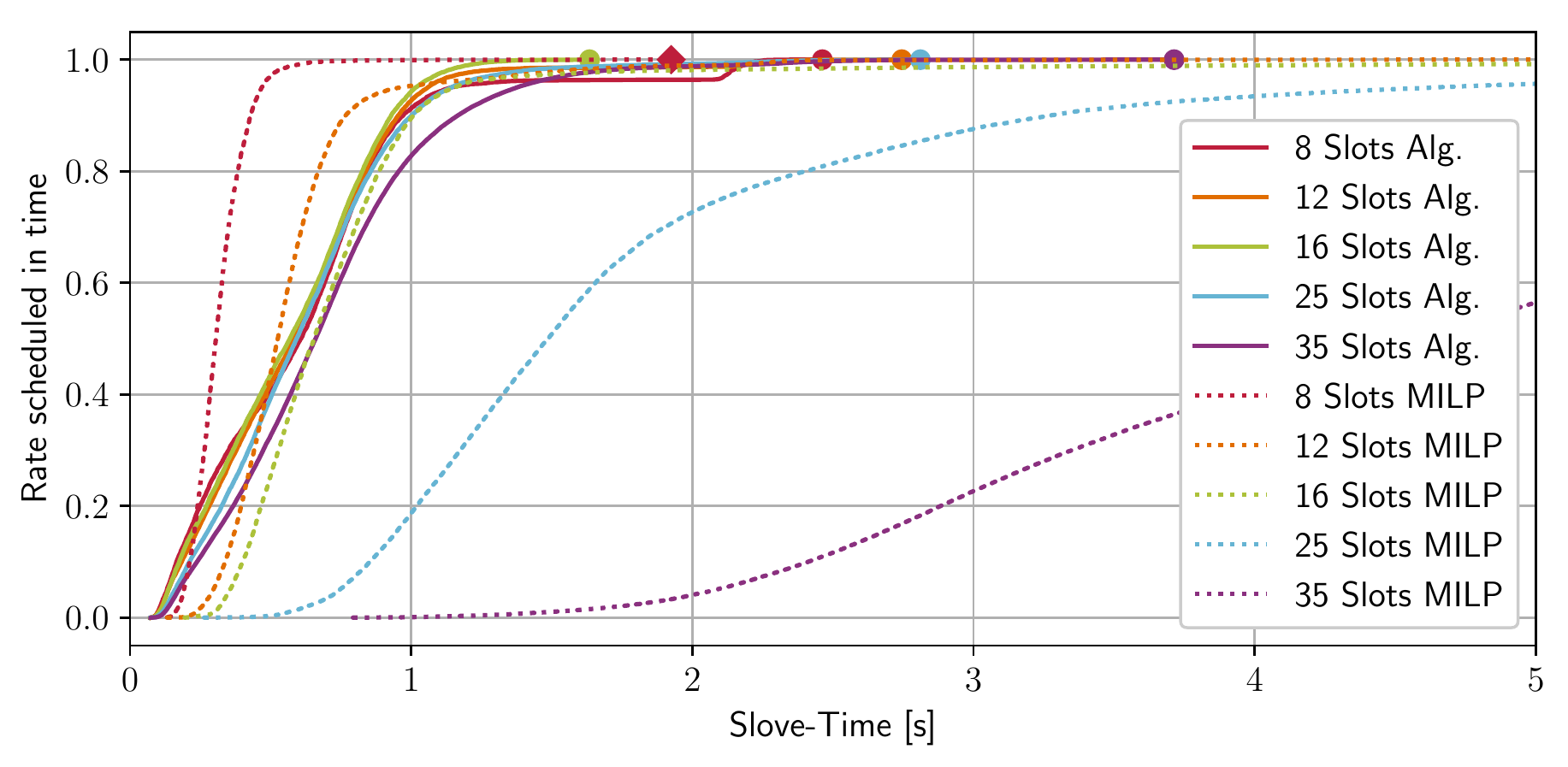}
\caption{\label{fig:compareSlotsTime} Influence of hyperperiod length on the schedulability, comparing the four algorithm modes and the \ac{milp}-model. For three jobs, nine dependencies and twelve nodes in mode [0, 0] \squeeze\vspace{-1mm}}
\end{figure}

\Cref{fig:compareSlotsTime} shows the comparison of the time needed to schedule the same tasksets with the proposed algorithm and the \ac{milp}-model.
The dotted lines represent the \ac{milp}-model and solid lines the heuristic algorithm, the maximum solve time is marked with a diamond for the \ac{milp}-model and a dot for the algorithm.
The results for the four different modes are very similar, therefore we only discuss the results of the \textit{time first shifting and age first} mode [0, 0]. 
\Cref{fig:compareSlotsTime} shows that the solve times of the algorithm are much closer to each other and, thus are much less influenced by the length of the hyperperiod.
With the vast majority of tasksets scheduled in less than one second, the algorithm is much more likely to finish scheduling in time even on embedded devices.
Nevertheless, for short hyperperiods the \ac{milp}-model is faster, that suggests that there is room for optimizations, at least in the implementation of the algorithm. 

\subsection{Influence of Taskset Parameters to Scheduling Success}
\label{sec:infl-tasks-param}

To determine under which conditions the algorithm is able to schedule what percentage of the tasksets, we used the same set of tasksets as in \Cref{sec:eval-compl-comp}.

\begin{figure}[hbt!]
\centering
\squeeze
\includegraphics[width=0.9\linewidth]{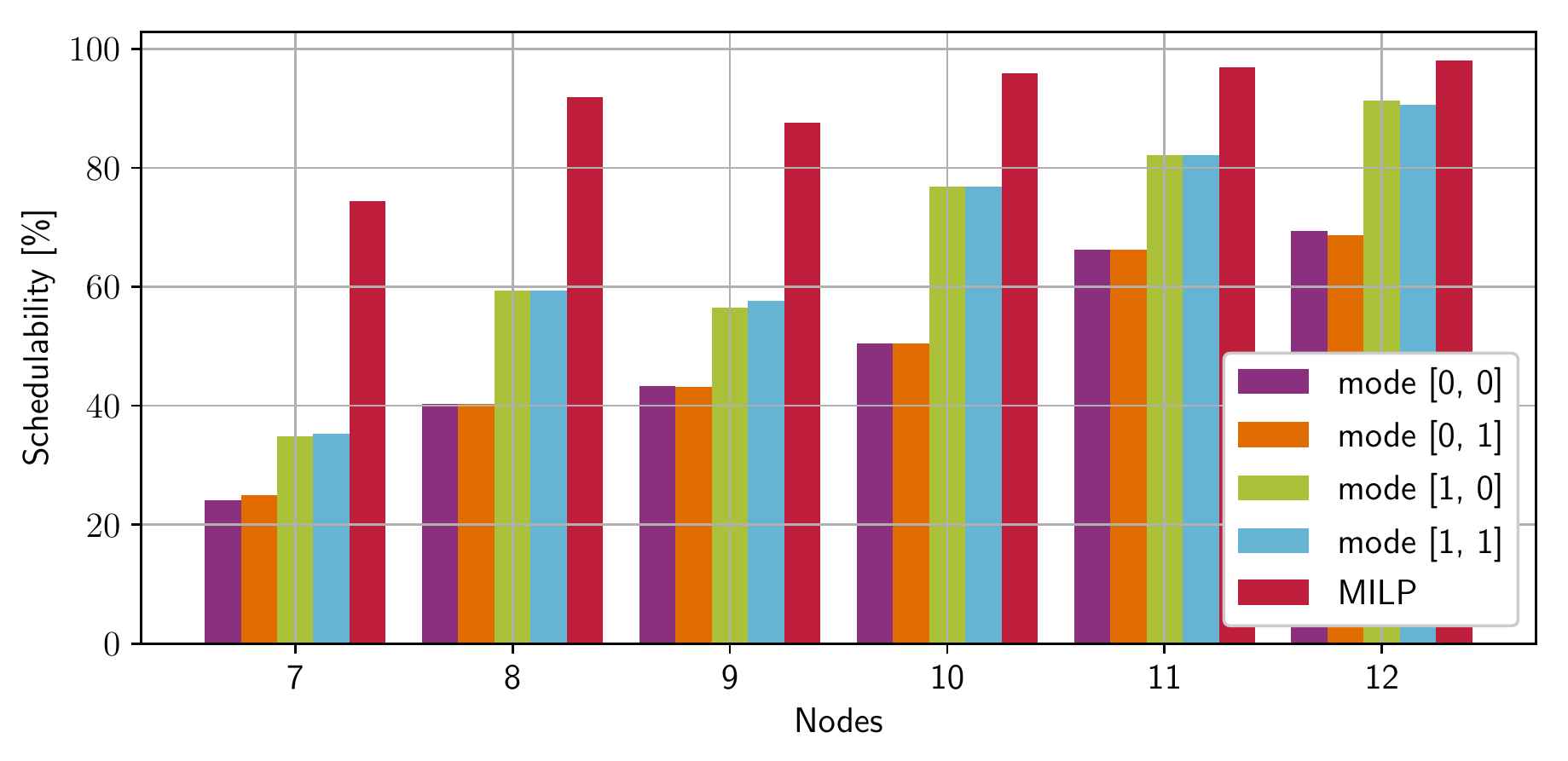}
\caption{\label{fig:compareNodes} Influence of \ac{tc}-size on the schedulability, comparing the four algorithm modes and the \ac{milp}-model. For three jobs, nine dependencies and 35 slots \squeeze}
\end{figure}

The biggest influence is caused by the number of nodes in a \ac{tc}, as shown in \Cref{fig:compareNodes}.
In general the results show that a taskset, with all the same parameters except for the number of nodes, is harder to schedule if less nodes are in the \ac{tc}.
This the due to the fact that in such tasksets there are more intersecting task which makes it more likely that a taskset is unschedulable.
Therefore, the algorithm mode should be compared to the \ac{milp}-model.
This comparison still shows that the algorithms performance deteriorates if fewer nodes are in a \ac{tc}.
The results of this evaluation confirm the assumption further that \textit{channel first shifting} is the superior mode.
They also give rise to the assumption that the algorithm is strongly effected by the number of dependencies.

\begin{figure}[hbt!]
\centering
\squeeze
\includegraphics[width=0.9\linewidth]{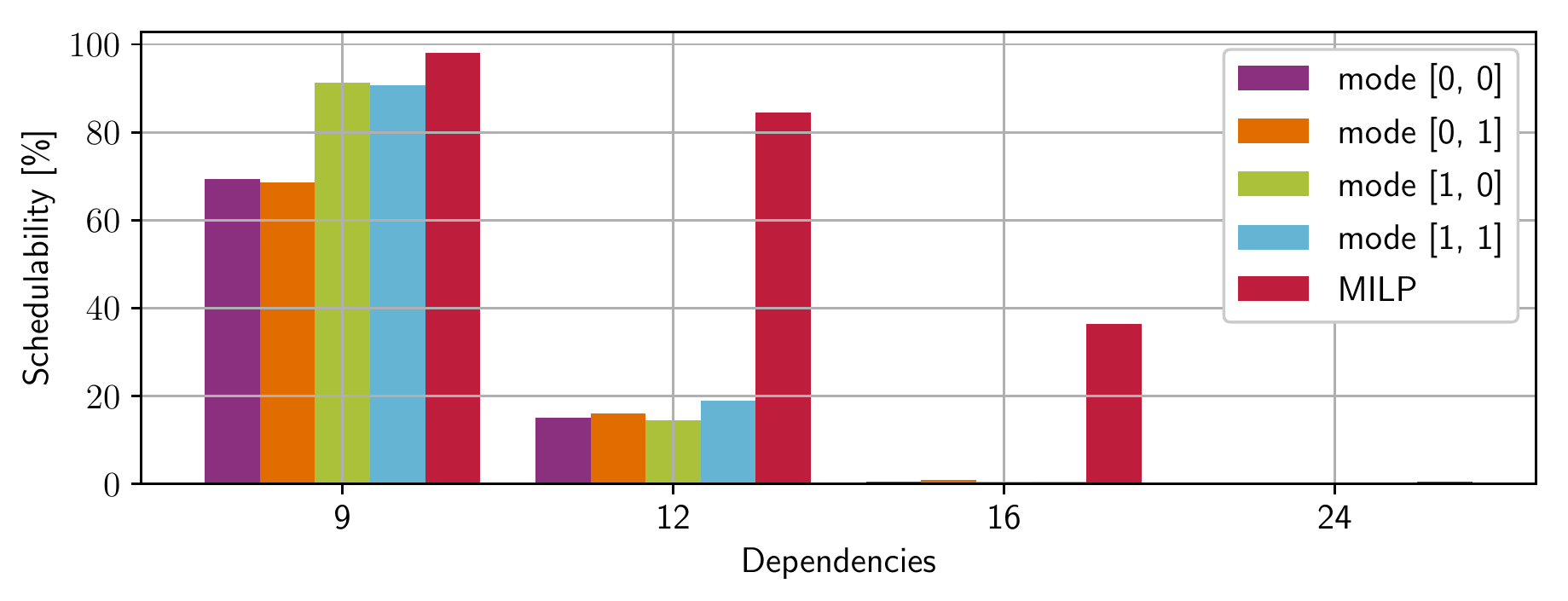}
\caption{\label{fig:compareDeps} Influence of the number of dependencies on the schedulability, comparing the four algorithm modes and the \ac{milp}-model. For three jobs, and 35 slots and twelve nodes \squeeze}
\end{figure}

To confirm the assumption we evaluated the performance of the algorithm on tasksets with the same parameters except for the number of dependencies.
As for the evaluation above, the algorithms performance needs to be compared with the \ac{milp}-model.
The results depicted in \Cref{fig:compareDeps} show a dramatic decline in the percentage of schedulable tasksets between nine and twelve dependencies.

Due to space limitations we are not able to show the influence the hyperperiod and number of jobs have.
In general the hyperperiod length does not have a significant influence, the number of jobs on the other hand has an influence.
For a constant number of tasks, a higher number of jobs is harder to schedule than a lower number.
This is due to the fact, that the algorithm tends to fill time-slots that are a multiple of the jobs periods first.
By having more jobs the chance that this slot in the ones in the jitter range are already occupied is higher.

\begin{figure}[hbt!]
\centering
\squeeze
\includegraphics[width=0.9\linewidth]{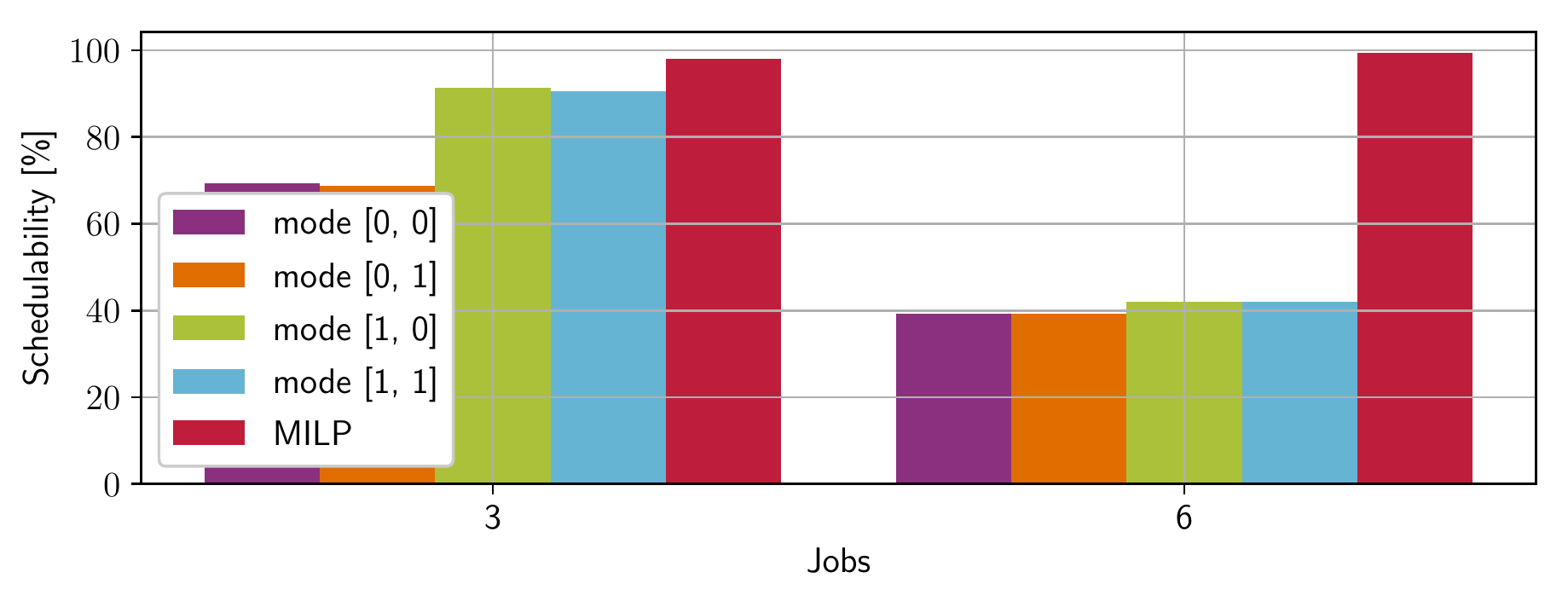}
\caption{\label{fig:compareJobs} Influence of the number of jobs on the schedulability, comparing the four algorithm modes and the \ac{milp}-model. For nine dependencies, and 35 slots and twelve nodes\squeeze}
\end{figure}

As the algorithm tries to schedule the leaf task of each job in the last slot of the subperiod, we assume that more jobs lead to a lower performance.
To confirm the assumption we evaluated tasksets with the same parameters but the varied number of jobs.
\Cref{fig:compareJobs} show that this assumption is true, it also shows that the performance of \textit{channel first shifting} declines more than the performance of \textit{time first shifting}.

All evaluations show that the \textit{channel first shifting} mode was superior to the \textit{time first shifting}, only in some cases the performance of both approaches were close to each other.
The evaluation also shows that the order in which the tasks are chosen to be scheduled does not make a noticeable difference in most cases and if there is a difference there is no pattern behind which of them performs better.

\subsection{Slot Allocation Probability}
\label{sec:slot-alloc-prob}
\begin{figure}[hbt]
\centering
\squeeze
\includegraphics[width=.8\linewidth]{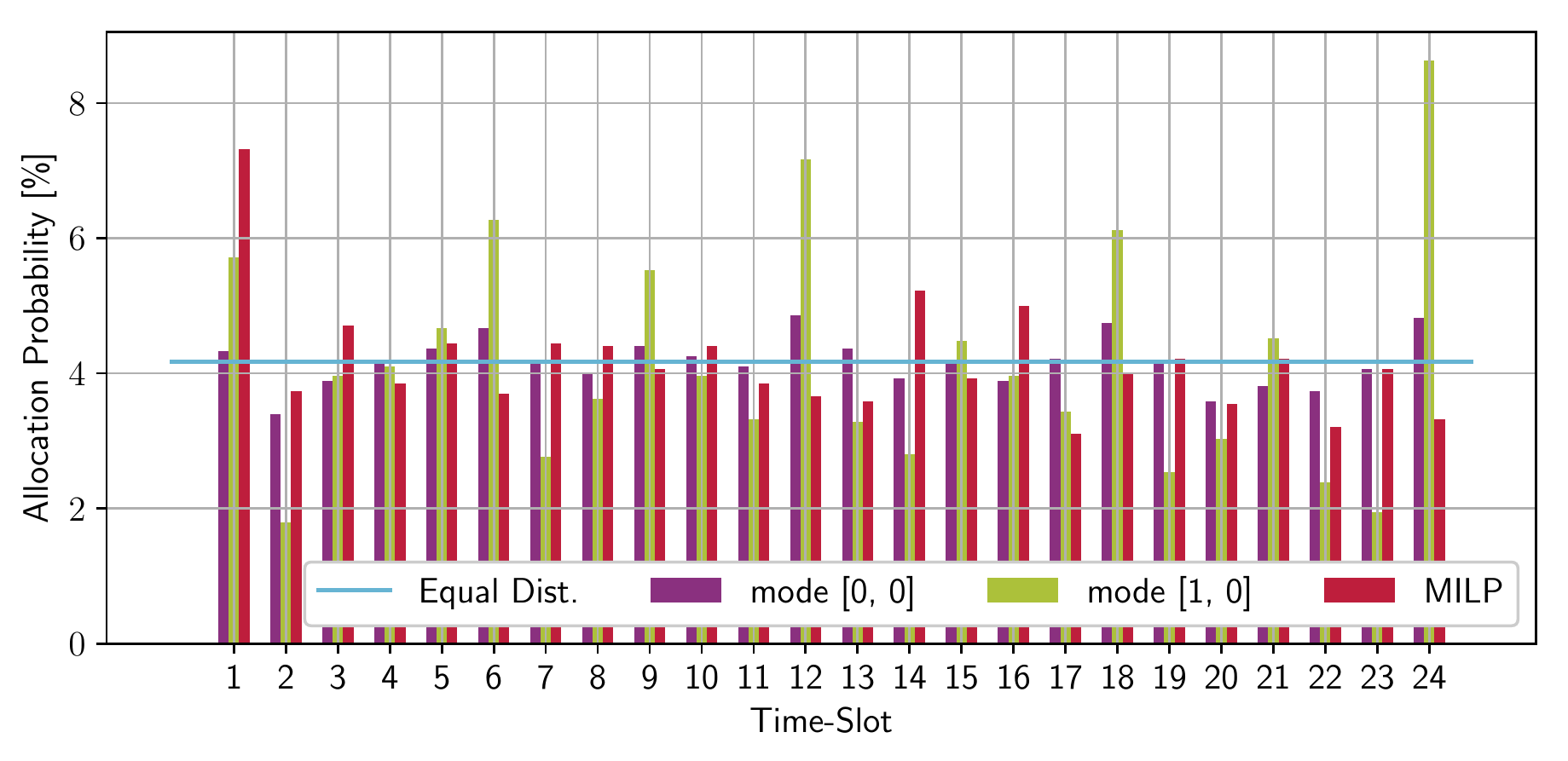}
\caption{\label{fig:compareDist} Comparison of the probability that a certain time-slot is allocated by the different scheduling approaches, for tasksets with hyperperiod length 24 \squeeze\squeeze}
\end{figure}

To investigate further why \textit{channel first shifting} suffers more from an increased job number than \textit{time first shifting} we compare the probability of each time-slot to be allocated for a task.
\Cref{fig:compareDist} shows these probabilities for both shifting approaches and the \ac{milp}-model, it also depicts the value for an equal distributions as a reference. 
As we discussed above the algorithm always tries to allocate the last time-slot in the subperiod of a task.
For the \textit{channel first shifting} this behavior is evident from the results, time-slots that are divisors of 24 have a higher probability to be allocated.
The higher the divisor is, the higher is the probability, as there are more jobs that share the end of its subperiod here.
For the \textit{time first shifting} this effect is still noticeable but it is mitigated.
The \ac{milp}-model does not show this effect, as it is not bound to the limitations in the heuristic, but it shows the tendency to allocate the first time-slot.

\subsection{Allocation Introduced Jitter}
\label{sec:alloc-intr-jitt}
\begin{figure}[ht]
  \centering
\squeeze
    \centering\includegraphics[width=.7\linewidth]{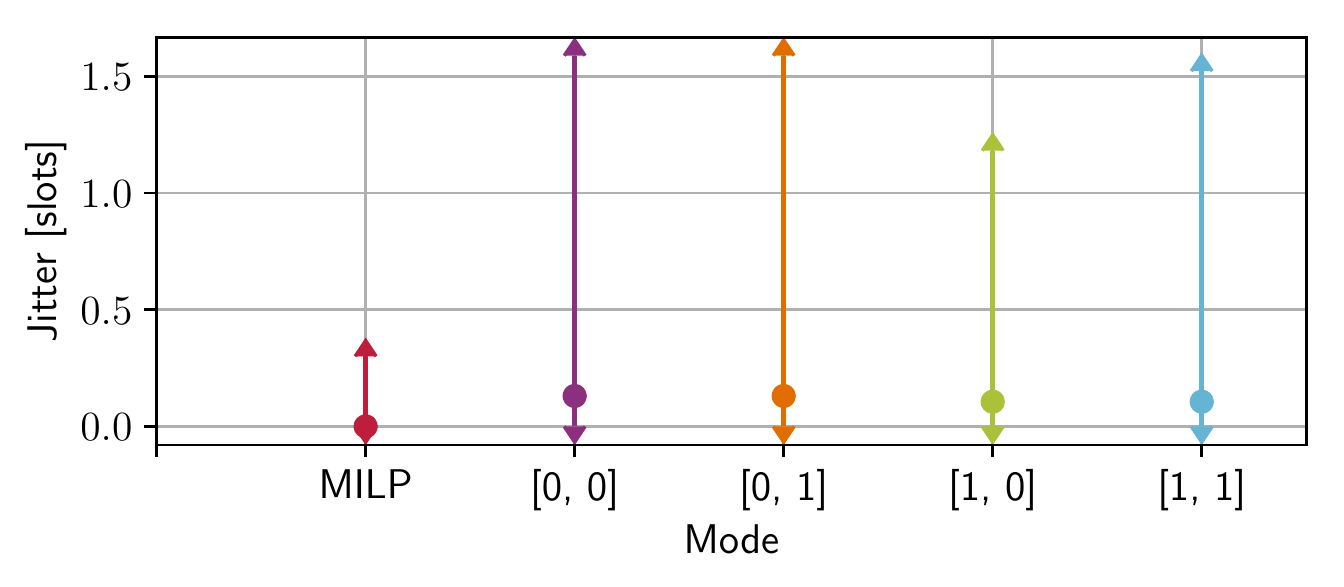}
  \caption{the mean, minimum and maximal jitter of the four algorithm modes in comparison to the \ac{milp}-model \label{fig:AllocSchedJitterCompare}\squeeze}
\end{figure}

The jitter a scheduling approach introduces is, after its ability to schedule tasksets, one of the most important performance factors of a real-time scheduling approach.
To evaluate this factor we compare the jitter introduced by the four modes of the proposed algorithm with the \ac{milp}-model in \Cref{fig:AllocSchedJitterCompare}.
It shows the minimum, mean and maximum jitter each mode introduces into a set of 56000 tasksets which are all scheduleable by all modes.
As in the previous evaluations the \textit{channel first shifting} outperforms the \textit{time first shifting}.
Especially with the \textit{ascending age} order the maximum jitter is lower.
The mean jitter on the other hand does not differ significantly over all modes.
The \ac{milp}-model with jitter optimization outperforms all algorithm modes, as it tries to find the minimal possible jitter.

\Cref{fig:AllocSchedJitterCompare} shows that the impact of the ordering is less significant than the impact of the shifting.
The impact of the ordering is so small that the lines are actually overlapping in the graph.

The results of this evaluation are very promising especially for an algorithm that is not actively reducing the jitter but only relies on its allocation methods.  

\subsection{Performance of Rescheduling}
\label{sec:perf-resch}
\begin{figure}[hbt!]
\centering
\squeeze
\includegraphics[width=.8\linewidth]{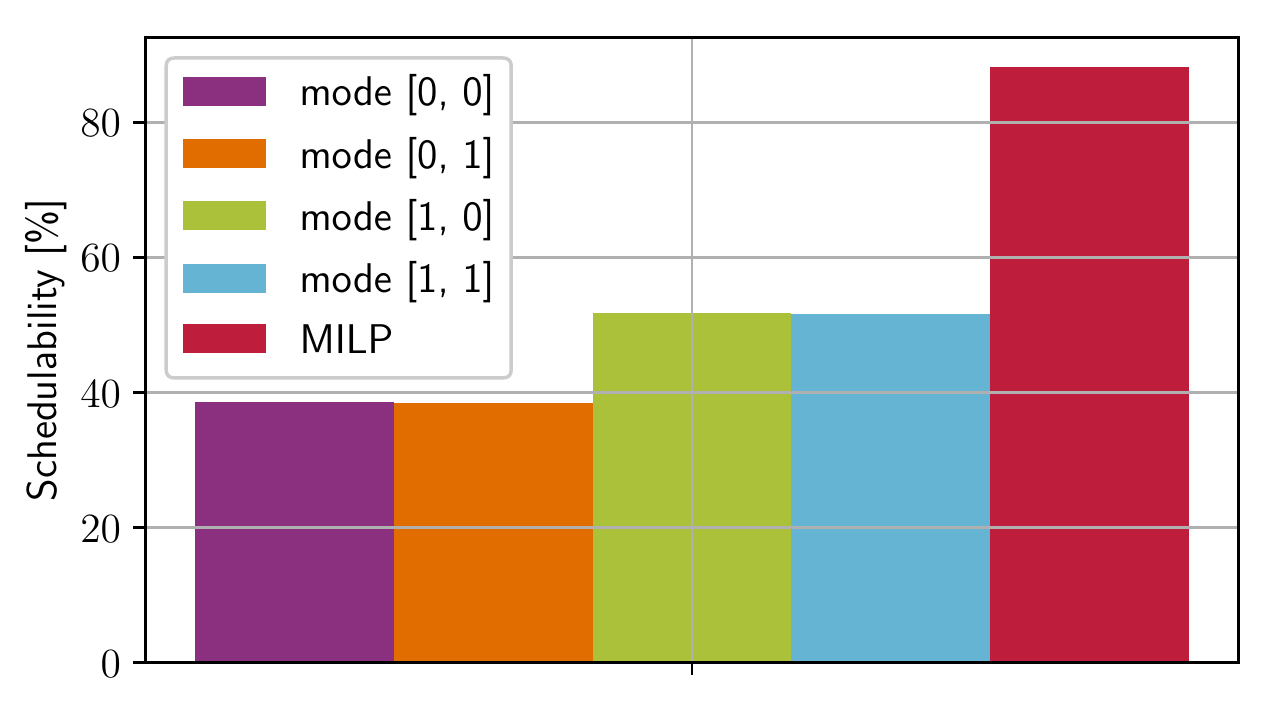}
\caption{\label{fig:compareReschedule} Comparison of the scheduling performance for taskset combinations  \squeeze\squeeze}
\end{figure}

To evaluate how good the different approaches are able to schedule combinations of tasksets we generated over two million combinations of the scheduleable tasksets.
From these we randomly chose a set of 250\,thousand combinations.
This set was scheduled by all four modes of the proposed algorithm and the \ac{milp}-model as a reference.
\Cref{fig:compareReschedule} depicts the percentage of combinations that were scheduleable by each mode.
As all evaluations above this shows that the \textit{channel first shifting} is superior to the \textit{time first shifting}.

The \textit{channel first shifting} managed to schedule over 50\,\% of the combinations.
This seems to be a quiet low percentage, but taking into consideration that we formed the combinations from tasksets with the same hyperperioid, job number and node number, the effect discussed in \Cref{sec:slot-alloc-prob} is amplified here.
As even more tasks need to be scheduled in the same time-slots, these combinations are the worst case combinations.
Together with the fact that our tasksets are quite dense, \Cref{fig:compareDist} shows that almost all time-slots are used at on least one channel, a success rate of 50\,\% is a promising result.

\section{Conclusion }
\label{sec:alloc-concl}
As cooperative \acp{cps} are emerging in real-time applications, wireless connections become more and more important.
These applications need scheduling algorithms which are able to adapt schedules to new network topologies and application requirements.
In \Cref{sec:AllocSchedProblem} we described how a cooperative \ac{cps} might be modeled and what the special challenges in these systems are.
We discussed the constraints a scheduling algorithm for such systems needs to fulfill in \Cref{sec:sched-constr-object}.
After a review of existing work in related areas we developed the \ac{milp}-model that generates schedules following the constraints discussed earlier.
As it can be time consuming to solve the proposed \ac{milp}-model, especially on embedded devices, we propose an algorithm to generate schedules which has a lower computation time for almost all cases.
To generate schedules that are easy to combine with other schedules we stated the hypothesis that two more sparse schedules are easier to combine than two dense schedules, even with the same hyperperiod and number of tasks.
We proof the hypothesis in \Cref{sec:adaptibilityHypothesis}.
In \Cref{sec:AllocSchedEuch} we discuss the proposed algorithm and its four different modes in detail.
The evaluations in \Cref{sec:scheduleAlgoEvaluation} compares these four modes to each other and to the \ac{milp}-model.
We show that the \textit{channel first shifting} is the preferable of the two shifting modes.
The order in which tasks are scheduled, on the other hand has no distinguishable influence on whether a taskset is scheduleable by the algorithm or not.

\bibliographystyle{abbrvnat}
\bibliography{ref}

\end{document}